
\magnification\magstep1
\hsize=16.5truecm
\vsize=22truecm
\baselineskip=14pt
\parindent=1em
\parskip=0.5ex
\voffset=0.truecm 
\hoffset=0cm 
\tolerance=1000
\def\eps{{\varepsilon}}
\def\d{{\rm d}}
\def\v{v_{j,k}}
\def\u{u_{j,k}}
\def\vv{v_j^{(n)}}
\def\uu{u_j^{(n)}}
\def\p#1#2{{\partial#1\over\partial#2}}
\def\pp#1#2{{\partial^2#1\over\partial#2^2}}
\def\frac#1#2{{\displaystyle#1\over\displaystyle#2}}
\def\T{\langle T\rangle}
\def\Tt{\langle T_0\rangle}
\def\Tb{\langle T_{\rm {b}}\rangle}
\def\inf{{{\rm inf}(j,\ell)}}
\def\sup{{{\rm sup}(j,\ell)}}
\def\Dpar{{\bf D}_\Vert}
\def\Dperp{{\bf D}_\perp}
\font\ttlfnt=cmr10 scaled 1500

\noindent {\ttlfnt Spontaneous symmetry breaking:
\noindent exact results for a biased random walk model of an exclusion process}
\bigskip
\noindent
C. Godr\`eche$^{1}$, J. M. Luck$^{2}$, M. R. Evans$^{3}$, D. Mukamel$^{4}$,
S. Sandow$^{4}$, and E. R. Speer$^{5}$

\bigskip
\noindent
$^{1}$Service de Physique de l'\'Etat Condens\'e,
Centre d'\'Etudes de Saclay, 91191 Gif-sur-Yvette cedex, France

\noindent
$^{2}$Service de Physique Th\'eorique,
Centre d'\'Etudes de Saclay, 91191 Gif-sur-Yvette cedex, France

\noindent
$^{3}$Theoretical Physics, The University of Oxford, 1 Keble road, Oxford,
OX1 3NP, U.K.

\noindent
$^{4}$Department of Physics of Complex Systems, The Weizmann Institute
of Science, Rehovot 76100, Israel

\noindent
$^{5}$Department of Mathematics, Rutgers University, New Brunswick, New Jersey
08903, U.S.A.

\bigskip
\noindent
{\bf Abstract.}
It has been recently suggested that a totally asymmetric exclusion process
with two species on an open chain
could exhibit spontaneous symmetry breaking
in some range of the parameters defining its dynamics.
The symmetry breaking is manifested by the existence of a phase in which
the densities of the two species are not equal.
In order to provide a more rigorous basis to these observations
we consider the limit of the process
when the rate at which particles leave the system
goes to zero.
In this limit the process reduces to a biased random walk in
the positive quarter plane, with specific boundary conditions.
The stationary probability measure of the position of the walker in
the plane is shown to be concentrated around two symmetrically
located points, one on each axis,
corresponding to the fact that the system is typically in one of
the two states of broken symmetry in the
exclusion process.
We compute the average time for the walker to traverse the quarter plane
from one axis to the other, which corresponds to the average time
separating two flips between states of broken symmetry in the
exclusion process.
This time is shown to diverge exponentially with the size of
the chain.

\vskip 2cm
\noindent
Submitted for publication to Journal of Physics A
\vskip 8 pt
\noindent

\vfill \eject

\vskip 14 pt
\noindent
{\bf 1 Introduction}
\vskip 12 pt

It is well known that
systems in thermal equilibrium do not exhibit spontaneous
symmetry breaking in one dimension at finite temperature,
provided that the interactions are short range
and that the local variable which describes the microscopic state of the
system takes only a finite number of possible values [1].
Common examples of such systems are Ising or Potts-like models.
On the other hand, when the local
variable takes one of an infinite number of possible values the model may
exhibit phase transitions and symmetry breaking in one dimension.
Examples are solid-on-solid models which are used to describe
wetting phenomena or unbinding transitions of a one dimensional interface
from an attractive wall [2].

Systems which are not in thermal equilibrium, but rather
evolve under some stochastic dynamics, are different.
In many cases these systems reach a steady state which does
not obey detailed balance.
The question is whether such a steady state of a one dimensional system
can break the symmetry of the dynamical rules under which the
system evolves.
A related
problem has been considered in the context of error correcting
computation algorithms, and an example of a one dimensional array of
probabilistic cellular automata which exhibits nonergodicity (as would a
model with spontaneous
symmetry breaking) has been constructed [3].
However, this example is rather complicated and not widely understood.

Recently a simple model was introduced [4, 5],
describing a totally asymmetric exclusion process for two species of particles
moving on an open chain of length $N$.
Each lattice site may be occupied
by either a positive $(+)$ or a negative $(-)$ particle, or by a hole
$(0)$.
Positive particles move to the right, negative particles to the left.
During an
infinitesimal time interval $\d t$ the following exchange events may take
place.
On the one hand, positive particles can exchange with adjacent holes to
their right, negative particles with adjacent holes to their left,
and positive particles with adjacent negative particles to their right,
at rates $1$, $1$, and $q$, respectively.
That is, between two adjacent sites one has
$$
+\, 0 \to 0\, + ,\qquad
0\, - \to -\, 0 ,\qquad
+\, - \to -\, + ,\qquad
\eqno (1.1)
$$
with probabilities $\d t$, $\d t$ and $q\d t$.
On the other hand, positive particles are
introduced at the left end with rate $\alpha$
and leave the system at the right end with rate $\beta$,
and conversely for negative particles.
Thus, at site 1
$$
0 \to + ,\qquad   - \to 0,
\eqno (1.2)
$$
with probabilities $\alpha \d t$ and $\beta \d t$ respectively;
at site $N$
$$
\eqalign{
0 \to - ,  \qquad + &\to 0,
}
\eqno (1.3)
$$
with probabilities $\alpha \d t$ and $\beta \d t$ respectively.
The dynamical rules of the model are symmetric under the simultaneous
interchange of the sign of particles and of left and right,
i.e., the dynamics of the
positive particles moving to the right is the same as that of the
negative particles moving to the left.
Therefore, as long as this symmetry is not spontaneously broken, the
densities and currents of the positive and negative particles are expected
to be equal.

The model has been studied in [4, 5]
both in a mean field approximation and by
simulation.
The mean field phase diagram exhibits a broken symmetry phase when the
rate $\beta$ is smaller than a critical value $\beta_c(\alpha, q)$;
in this phase the
currents and the densities of the positive and negative particles are unequal.
Numerical simulations of the model give similar results: the system
evolves to one of two possible long lived states.
These two states are
related by the symmetry of simultaneous sign of particles
and left-right interchange.
In one of them the current and density of positive particles are larger
than those of the negative particles, and in the other these inequalities
are reversed.
Outside of small regions near the boundaries, the states appear to
be spatially uniform, and in the $N\to\infty$ limit
they become true steady states, hereafter also referred to as {\it phases},
to be identified with those of
the infinite system studied, in the case $q=1$, in [6].
For any finite size
$N$ the system flips between these two possible states as it evolves,
resulting in an overall symmetric state.
This dynamics is characterized by
a time scale $\langle T\rangle$, the average time between two consecutive
flips, which diverges in the $N\to\infty$ limit.
In [4, 5] it
has been suggested, using numerical
simulations and some simple considerations, that $\T$ grows exponentially
with $N$.
The flipping process itself takes place over a time interval much less than
$\langle T\rangle$.

To summarize, the two-species exclusion process studied in
[4, 5] is known,
at least on a heuristic basis,
to exhibit
the following two properties
(when $\beta<\beta_c$):

\noindent
(i) the system spends most of its time in one of the two symmetry related
long lived states described above,
each corresponding, in the $N\to \infty$ limit,
to a spatially homogeneous phase over the entire system,

\noindent
(ii) the average time between two consecutive flips,
i.e.,  the average lifetime of a long lived state,
is exponentially increasing with
the size of the system:
$$
\T\sim\exp(N\mu)
.\eqno(1.4)
$$
The positive constant $\mu$ in (1.4), hereafter referred to as the {\it
mass},
is the analogue in the present situation of the barrier height, or
activation energy, per site, characteristic of the law of Arrhenius
for systems at equilibrium.

These two properties, in analogy with the situation
which prevails in the case of the two dimensional Ising model
below the critical temperature,
lead us to say that the model exhibits spontaneous symmetry breaking.
The analogy between the two models may be pursued.
The escape rate $\beta$ in the exclusion process plays the role of
temperature, $\alpha$ and $q$ being kept fixed.
In particular, density fluctuations increase with $\beta$,
and a transition from a broken symmetry phase to a symmetric phase
is found when $\beta=\beta_c$.
Finally changing the input and output rates at the boundaries
to produce an asymmetry between positive and negative particles
is akin to introducing a magnetic field in the Ising model
(see the conclusion).

The aim of the present paper is to demonstrate
the existence of spontaneous symmetry breaking
in the $\beta\to 0$ limit of the two-species exclusion model,
by showing that the two properties mentioned
above hold.
In this limit the model can be
mapped onto a much simpler {\it toy model}, for which the set of available
configurations is restricted to those composed of three blocks, with
$j$ negative particles on the left, $\nu$ holes in the middle
and $k$ positive particles on the right, with $N=j+k+\nu$
(see eq.~(2.1)).
The dynamics of these configurations
may be represented by a biased random walk in the positive quarter plane
$(j,k)$,
with boundary conditions to be defined below.
We demonstrate that the toy model
flips between two long lived states.
These states coincide, except near the
boundaries, with one or the other pure phase of the system, i.e.,
typically the larger of the $(-)$ and $(+)$ blocks fills the system
except for a region whose size does not grow with $N$:
with high probability, $\max(j,k)=N-O(1)$.
We are able to compute the average time between two consecutive flips,
and find an exponential growth with the system size
according to eq.~(1.4).

Let us finally emphasize that both properties:

\noindent
(i) existence of pure phases (related by symmetry),

\noindent
(ii) divergence of the time scale
$\langle T\rangle$ with the system size,

\noindent are necessary in order to claim spontaneous symmetry breaking,
neither of them taken separately being sufficient.

This may be illustrated by the following case.
In the one-species exclusion process on an open chain, with input rate
$\alpha$ and output rate $\beta$, studied in
[7-10],
the typical configuration on the phase coexistence line $0<\alpha=\beta<1/2$
consists of two regions of size $O(N)$, one in each phase
(one of low density $\alpha$, the other of high density $1-\alpha$),
separated by a {\it shock}.
The shock diffuses, hence its position is uniformly distributed over the
chain in the steady state, and there is a characteristic time of
order $N^2$ for the shock to traverse the system and hence for the system
to change from one pure phase to another,
without ever staying in one of them.
In other words, there are no pure long lived states in this case;
the steady state is a mixture.
In conclusion, though the time scale $\T$ necessary for the system to pass
from one phase to the other (without staying in one of them) diverges
with the size, no spontaneous symmetry breaking occurs in this case.
Note that the $N^2$ divergence of $\T$, instead of an exponential one,
comes together with the absence of pure phases.
This situation is presumably generic.
We will come back to this example in section 5 since
it corresponds to a limiting case of the two-species process.

This paper is organized as follows.
In section~2 we present simple physical
arguments suggesting that, in the $\beta\to 0$ limit,
the dynamics of the exclusion process
may be mapped onto that of the toy model.
This correspondence is made rigorous in Appendix~A.
Section~3 provides a description of the toy model
as a biased random walk in the positive quarter plane.
The equations for the stationary probability
are given in section~4, as well as those for the
average time between two flips.
The latter involves
two intermediate functions $\u$ and $\v$, which have simple
probabilistic interpretations.
Appendix B provides a convenient way to compute all these quantities
numerically.
Section~5 is devoted to the study of two simple limiting cases of the
random walk which are helpful to understand the general case.
The analysis
of the general case is given in section~6, where the asymptotic expressions
of the stationary probability  measure and of
the average time between two flips are derived.
We thus show that the
steady state in the toy model is concentrated on configurations in which
the system is almost entirely in a single phase.
The analytic expression of the mass $\mu$ is given.
In section~7 we determine the scaling behavior of $\u$, $\v$ and
the average time between two flips in the continuum limit,
corresponding to a small bias and large
distances.
A summary and a discussion are given in section~8.

\vskip 14 pt
\noindent
{\bf 2 Definition of the toy model}
\vskip 12 pt

In this section we show by means of simple physical arguments that in the limit
where the rate $\beta$ at which particles leave the system is small
compared to the other rates defining the
dynamics---$\alpha$ (particle input), $q$ (particles interchange) and
$1$~(particle-hole exchange)---the exclusion model reviewed in the
Introduction may be mapped onto a simpler model, which we term the {\it toy
model}.
A rigorous proof of the equivalence of the two models is provided in Appendix
A.

The time scale for the toy model is determined by the escape rate $\beta$,
and we therefore introduce a rescaled time $\tau=2\beta t$.
In the $\beta \to 0$ limit
the only relevant configurations are
those composed of three blocks, with negative particles to the left, holes
in the middle and positive particles to the right:
a typical configuration is of the form
$$
\overbrace{\underbrace{-\cdots -}_{\displaystyle j}\,
\underbrace{{\vphantom{+}}0\cdots 0}_{\displaystyle \nu}\,
\underbrace{+\cdots +}_{\displaystyle k}}^{\displaystyle N}.
\eqno(2.1)
$$
All other configurations may be neglected.
This can be understood as follows.

First, since $\beta$ is small, outputs of particles are rare on the time
scale of the exclusion model.
Any configuration different from those above
will therefore rearrange itself---quickly, on the $\tau$ time scale---until
all the particles are waiting to exit. Suppose now that a
particle exits.
Either it will be replaced by a particle of the opposite
sign, which will quickly traverse the system until it reaches the block of
particles of its sign, or the hole thus created will itself travel through
the system to join the block of holes in the middle.
On the $\tau$ time
scale, then, the dynamics of the system involves transitions between
various three-block configurations.
These transitions may be described only in terms of the events taking
place on the boundaries.
Note that if the particle
which exits is the last of its type, then the system will
quickly fill with particles of the other type, leaving a configuration
with only one block.
Moreover, should the system empty completely, particles will
quickly enter and rearrange to the form (2.1).
Thus all configurations with $j=0$ or $k=0$,
except for those with $k=N$ or $j=N$, respectively,
may be neglected in the $\beta\to 0$ limit.

Let us now give a quantitative description of the dynamics in this
configuration space.
We denote by $(j,k)$ a three-block configuration,
according to eq.~(2.1).
The integers $j$ and $k$  may take the
values $0, 1,\ldots, N$, with the restriction that $j+k \le N$.
Configurations $(0,k)$ with $0\le k<N$ and $(j,0)$ with $0\le j<N$ do
not occur.

Consider now what may happen during a time interval $\Delta\tau$ which is
short on the $\tau$ time scale but long on the $t$ scale:
$\beta\ll\Delta\tau\ll1$.
We can ignore the possibility that more
than one particle exits in such an interval.
If $j\ge 1$,
then there is a small probability $(1/2)\Delta\tau=\beta\Delta t$
that the negative particle at
site 1 leaves the system during this time interval.
Suppose that this happens.

\noindent
(i) If $j >1$, then immediately (on the $\tau$ scale) either the
hole thus created is filled by a positive particle, or the negative
particle on site 2 exchanges position with the hole; the relative
probabilities of these events are $\alpha/(1+\alpha)$ and
$1/(1+\alpha)$.
Either event will be followed by a fast reordering of
the particles, resulting in the configuration $(j-1,k+1)$ or $(j-1,k)$,
respectively, unless $j=N$, when in the second case an additional
negative particle will enter once the hole reaches the right end of the
interval, resulting in the final configuration $(N,0)$, that is, in no net
change.

\noindent
(ii) If $j=1$, then
when the negative particle at site $1$ leaves the system,
the negative block disappears altogether.
In this case positive particles are quickly introduced into
the system at the left end
and the system becomes filled with positive particles.
The resulting configuration is $(0,N)$.

\noindent Similar processes take place at the
right end when a positive particle leaves the system.

To summarize these results we introduce rescaled rates
$$
a={1\over 2(1+\alpha)}, \quad b={\alpha\over 2(1+\alpha)},
\eqno(2.2)
$$
which satisfy
$$
a+b=1/2.
\eqno(2.3)
$$
Then during the time
interval $\d\tau$ the following stochastic
changes take place in the system:
$$
\left\{\!\!\!\!\!\!\matrix{
&(j, k)&\to &(j-1,k+1)\hfill&{\rm if}\;j>1,\quad\hfill
&{\rm with~probability}\quad b\,\d\tau,\hfill\cr\cr
&(j, k)&\to &(j-1,k)\hfill&{\rm if}\;N>j>1,\quad\hfill
&{\rm with~probability}\quad a\,\d\tau,\hfill\cr\cr
&(1,k)&\to &(0,N),\hfill&&{\rm with~probability}\quad
(1/2)\,\d\tau,\hfill\cr\cr
&(j, k)&\to &(j+1,k-1)\hfill&{\rm if}\;k>1,\quad\hfill
&{\rm with~probability}\quad b\,\d\tau,\hfill\cr\cr
&(j, k)&\to &(j,k-1)\hfill&{\rm if}\;N>k>1,\quad\hfill
&{\rm with~probability}\quad a\,\d\tau,\hfill\cr\cr
&(j,1)&\to &(N,0),\hfill&&{\rm with~probability}\quad
(1/2)\,\d\tau.\hfill\cr}\right.
\eqno(2.4)
$$
Let us emphasize again that states $(j,0)$ with $j<N$ and $(0,k)$
with $k<N$ do not occur.

These rules completely specify the toy model.
They describe a random walk in the $(j, k)$ plane, biased
towards the south-west direction, that we study hereafter.

\vskip 14 pt
\noindent
{\bf 3 The toy model as a random walk}
\vskip 12 pt

The rest of this paper is devoted to a quantitative analysis of
the random walk defined by the dynamical rules (2.4) of the toy model.
For the sake of clarity we summarize its definition first.

A random walker hops from site to site on a two dimensional lattice,
restricted to the positive quarter plane.
The position of the walker is denoted by $(j,k)$.
During the time interval $\d \tau$ it hops
from $(j,k)$ to the south $(j,k-1)$ or to the west $(j-1,k)$,
each with probability $a\,\d \tau$.
It may also hop to the north-west
$(j-1,k+1)$ or to the south-east $(j+1,k-1)$
each with probability $b\,\d \tau$ ($a+b=1/2$).
When the walker reaches the $j$  or the $k$ axis
it restarts instantly at site $(N,0)$ or $(0,N)$, respectively.
The random walker is thus subjected to non local boundary conditions.
For short we will say that the axes are {\it bouncing} boundaries.

Let us determine the velocity vector and the diffusion tensor
of this random walk.
Consider the probability $p_{j,k}(\tau)$ to find the walker
on site $(j,k)$ at time $\tau$.
Its evolution in time is given by the following master equation,
obtained by conditioning on the last step of the walker
$$
\frac{\d}{\d\tau}p_{j,k}(\tau)=a\big(p_{j+1,k}(\tau)+p_{j,k+1}(\tau)\big)
+b\big(p_{j+1,k-1}(\tau)+p_{j-1,k+1}(\tau)\big)-p_{j,k}(\tau)
,\eqno(3.1)
$$
leaving aside boundary conditions.
This equation may be rewritten as
$$
\frac{\d}{\d\tau}p({\bf x},\tau) =
\sum_ia_i\big[p({\bf x}-{\bf e}_i,\tau)-p({\bf x},\tau)\big],
\eqno(3.2)
$$
with ${\bf x}=(j,k)$, and where the ${\bf e}_i$ ($i=1,\dots 4$)
correspond to the four possible moves.
To the south: ${\bf e}_1=(0,-1)$,
to the west: ${\bf e}_2=(-1,0)$,
to the north-west: ${\bf e}_3=(-1,1)$,
to the south-east: ${\bf e}_4=(1,-1)$,
with rates $a_1=a_2=a$, and  $a_3=a_4=b$ respectively.

The velocity vector reads [11, 12]
$$
{\bf V}=\lim_{\tau\to\infty}{1\over\tau}\langle{\bf x}(\tau)\rangle
=\sum_i a_i {\bf e}_i,
\eqno(3.3)
$$
while the components of the diffusion tensor ${\bf D}$ are given by
$$
{\bf D}_{\mu\nu}=\lim_{\tau\to\infty}{1\over 2\tau}
\big\{\langle x_\mu(\tau)x_\nu(\tau)\rangle
-\langle x_\mu(\tau)\rangle\langle x_\nu(\tau)\rangle\big\}
={1\over 2}\sum_ia_i\,(e_i)_\mu(e_i)_\nu.
\eqno(3.4)
$$
In the present case one finds
$$
{\bf V}=\pmatrix{-a\cr -a},
\eqno(3.5)
$$
and
$$
{\bf D}={1\over 2}\pmatrix{1-a&2a-1\cr 2a-1&1-a}.
\eqno(3.6)
$$
The eigenvalues of this matrix yield the two diffusion coefficients
$$
{\bf D}_{\perp}={2-3 a\over 2},\quad {\bf D}_{\parallel}={a\over 2},
\eqno(3.7)
$$
where perpendicular and parallel refer to the south-west direction of the
velocity (or bias).
Note that the velocity and the parallel diffusion coefficient vanish as
$a\to0$.

The toy model captures the essence of the original exclusion process, when
$\beta$ is small; in particular the mechanism leading
to spontaneous symmetry breaking has an intuitive explanation
in terms of the random walk.
Let us first review this mechanism for the exclusion process.
Suppose the system is in the state of high density of $(+)$ particles.
In this phase there is a
low flux of negative particles and holes entering the system at the right end
and leaving at the left end.
As the system evolves, a block of negative
particles may temporarily be formed at the left end due to some
fluctuation.
As a consequence holes entering the system become trapped
between the positive and negative regions.
Usually, the block at the left
end leaves the system after some time and the system relaxes to the all $(+)$
state.
However, if the block persists long enough, the system may be
filled with holes, and thus has a chance of flipping to the all $(-)$ state.

This mechanism has also a clear interpretation in the framework
of the biased random walk of the toy model.
Consider the
walker starting at $(0,N)$, i.e. in the all $(+)$ state.
With large probability the walker will not go far east against the bias.
It will therefore soon hit the $k$ axis and bounce to its starting point
$(0,N)$.
However, with a small probability, vanishing as $N\to\infty$,
it may reach the other
end of the system $(N,0)$.
It is thus clear that the walker spends most of its time moving near one of the
two endpoints of the system, occasionally traversing to the other end.
Therefore the stationary probability measure
of the walker is concentrated around both endpoints $(0,N)$ and
$(N,0)$.
The average time for the
walker to move from one end of the system to the other is the
average time between two flips in the toy model.
This time is denoted hereafter by $\Tt$.
It is related to the average time between two flips of the original exclusion
process in the $\beta\to0$ limit by
$$
\T\approx{\Tt\over 2\beta}\quad(\beta\to 0).
\eqno(3.8)
$$

The stationary probability measure as well as the average time
between two flips in the toy model are studied in the next sections.

\vskip 14 pt
\noindent {\bf 4 Equations for the stationary probability and
for the average time between two flips}
\vskip 12 pt

Two questions are to be answered:

\noindent
(i) What is the probability $p_{j,k}$
for the walker to be at site $(j,k)$
in the steady state (i.e. the stationary probability measure)?

\noindent
(ii) What is the average time $D_{j,k}$ for the walker, starting from
site $(j,k)$,
to reach any site on the $j$ axis for the first time?
The average time between two flips in the toy model is simply
$$
\Tt=D_{0,N}.
\eqno(4.1)
$$

In this section we establish the equations fulfilled by these
two quantities.
Their solutions, mostly in the asymptotic $N\to \infty$ regime of interest,
are presented in the next sections.

\bigskip
\noindent
{\it 4.1 The stationary probability measure}

In the steady state,
the equation for the probability $p_{j,k}$
to find the walker on site $(j,k)$ is
$$
p_{j,k}= a\left(p_{j,k+1}+p_{j+1,k}\right)
+b\left(p_{j+1,k-1}+p_{j-1,k+1}\right)
\eqno(4.2)
$$
as a simple consequence of eq.~(3.1),
with boundary conditions
$$
\eqalign{
p_{j,0}&=0 \qquad (j\ne N)\cr
p_{0,k}&=0 \qquad (k\ne N)\cr
p_{j,k}&=0 \qquad (j+k> N).\cr
}
\eqno(4.3)
$$
The first two conditions are the only relevant consequences
of the bouncing property of the axes.

The stationary probability measure obeys the symmetry property
$$
p_{j,k}=p_{k,j}.
\eqno(4.4)
$$
In addition the following normalization condition is imposed
$$
\sum_{j,k} p_{j,k}=1.
\eqno(4.5)
$$

\bigskip
\noindent
{\it 4.2 The average time between two flips}

$D_{j,k}$, as defined above, is the first-passage time through the $j$ axis.
It is therefore convenient to consider this axis as absorbing
(the $k$ axis is still bouncing).
Conditioning on the first step of the walker,
during the infinitesimal time interval $\d \tau$, yields
$$
\eqalign{
D_{j,k}&
=a \d \tau(D_{j,k-1}+\d \tau)
+a \d \tau(D_{j-1,k}+\d \tau)\cr
&+b \d \tau(D_{j+1,k-1}+\d \tau)
+b \d \tau(D_{j-1,k+1}+\d \tau)\cr
&+[1-(2a +2 b) \d \tau](D_{j,k}+\d \tau),
}
\eqno(4.6)
$$
hence
$$
D_{j,k}
=1+a(D_{j,k-1}+D_{j-1,k})+b(D_{j+1,k-1}+D_{j-1,k+1}),
\eqno(4.7)
$$
where $j$ and $k$ $=1,\ldots,N$.
The boundary conditions read
$$
\eqalignno{
D_{0,k}&=D_{0,N}&(4.8{\rm a})\cr
D_{j,0}&=0 &(4.8{\rm b})\cr
b D_{0,N}&=b D_{1,N-1}+1.&(4.8{\rm c})
}
$$
The first equation reflects the non local bouncing property of the $k$ axis,
while the second one expresses that the $j$ axis is absorbing.
The last equation comes from the fact that, starting from
site $(0,N)$, the walker has probability $b\,\d\tau$
to hop to site $(1,N-1)$,
and the complementary probability $1-b\,\d\tau$ to stay on this site.
Note that the recursion relations found for the $D_{j,k}$, eqs.~(4.7),
are the same as those for a discrete-time process [13],
since the hopping rates of the walker add to 1.

In order to unravel the cumbersome condition eq.~(4.8a) imposed
by the bouncing boundary, it is convenient to introduce two functions
$\u$ and $\v$ satisfying
$$
\eqalignno{
u_{j,k}&=1+a(u_{j,k-1}+u_{j-1,k})+b(u_{j+1,k-1}+u_{j-1,k+1})&(4.9{\rm a})\cr
v_{j,k}&=a(v_{j,k-1}+v_{j-1,k})+b(v_{j+1,k-1}+v_{j-1,k+1}),&(4.9{\rm b})\cr
}
$$
with the boundary conditions
$$
\eqalignno{
u_{0,k}&=0,\quad v_{0,k}=1&(4.10{\rm a})\cr
u_{j,0}&=0,\quad v_{j,0}=0.&(4.10{\rm b})\cr
}
$$
$\u$ and $\v$ obey the symmetry properties
$$
u_{j,k}=u_{k,j},\quad v_{j,k}+v_{k,j}=1
.\eqno(4.11)
$$
Then the solution to eqs.~(4.7) and (4.8a-c) is given by
$$
D_{j,k}=u_{j,k}+D_{0,N}\,v_{j,k}
,\eqno(4.12)
$$
and in particular by
$$
D_{0,N}={u_{1,N-1}+1/b\over 1-v_{1,N-1}}
.\eqno(4.13)
$$

We are actually interested in determining $D_{0,N}$
in the asymptotic regime, physically relevant, where $N$ is large.
As shown by eq.~(4.13) two independent sets of equations
for $u_{j,k}$ and $v_{j,k}$ have to be solved separately.
This problem is studied in the next sections.

We conclude by giving
simple interpretations to the quantities met in
this section and to the relations between them, in particular to
the key eq.~(4.13).

\medskip
\noindent (i) As can be deduced by considering the equations
fulfilled by $u_{j,k}$, this quantity represents the
average time for the walker starting from $(j,k)$
to reach either of the axes for the first time.
Similarly $v_{j,k}$ is the
probability for the walker starting from $(j,k)$ to reach the
$k$ axis without hitting the $j$ axis.
Generalizing the language of the classical {\it Gambler's ruin} problem,
$u_{j,k}$ is the {\it duration of the game},
and $v_{j,k}$ the {\it probability of ruin}: the walker is ruined when
reaching the $k$ axis, whereas it wins when reaching the $j$ axis.
The game stops in both cases.
When considering the quantities $\u$ and $\v$ it is therefore convenient
to consider both axes as absorbing.

Hence equation~(4.13) also gets a simple interpretation.
Its denominator is the probability for the walker, starting from
$(1, N-1)$ to reach the $j$ axis, that we may name the {\it probability of
success}.
In the numerator, $1/b$ is the average time for the walker
starting from $(0,N)$ to reach site $(1,N-1)$,
while $u_{1,N-1}$ is the average time, starting from $(1,N-1)$,
to reach either of the axes.
Altogether the numerator of (4.13) represents
the average duration of an {\it attempt}
for the walker starting from $(0,N)$:
once the walker has reached site $(1,N-1)$,
either it succeeds with probability $1-v_{1,N-1}$
in traversing the quarter plane, or it fails with probability $v_{1,N-1}$,
returning to its starting point
(see also eq.~(4.19) below).
Alternatively this duration represents the average time between
two {\it bounces} on either of the axes.
Denoting this duration by $\Tb$, we have
$$
\Tb={1\over b}+u_{1,N-1},
\eqno(4.14)
$$
and (4.13) reads, using (4.1)
$$
\Tt={\Tb\over 1-v_{1,N-1}}.
\eqno(4.15)
$$
The probability of success is therefore the ratio of the
frequency of flips (or success) to the frequency of bounces
(or attempts).

\medskip
\noindent (ii)
The average time between
two bounces $\Tb$
can be independently evaluated as follows.
Just after a bounce the walker either passes through
the bond from $(0,N)$ to $(1,N-1)$,
or through the bond from $(N,0)$ to $(N-1,1)$.
Hence $\Tb=1/(2J)$,
where $J =b p_{0,N}$ is the probability current through either of the two
bonds.
Thus
$$
\Tb=\frac{1}{2bp_{0,N}}.
\eqno(4.16)
$$

Comparing eqs.~(4.14) and (4.16) yields
$$
{1\over 2\, p_{0,N}}=1+b\, u_{1,N-1}
\eqno(4.17)
$$
which will be used hereafter.

\medskip
\noindent (iii)
Finally, an alternative interpretation of (4.15) is provided
by the following considerations.
The average number of attempts of the walker until the first success is
$$
\langle m\rangle=\sum_{m\ge0} m\, (v_{1,N-1})^m (1-v_{1,N-1})
={v_{1,N-1}\over 1-v_{1,N-1}}
.\eqno(4.18)
$$
Denote by $u^{(j)}_{1,N-1}$ the average time
for a walker starting from $(1,N-1)$ to touch the $j$ axis
without touching the $k$ axis; $u^{(k)}_{1,N-1}$ is defined
similarly, by exchanging the roles of $j$ and $k$.
Then
$$
\Tt=\langle m\rangle \left({1\over b}+u^{(k)}_{1,N-1}\right)
+\left({1\over b}+u^{(j)}_{1,N-1}\right)
,\eqno(4.19)
$$
where the first parenthesis represents the average duration of
a failure, the walker starting from $(0,N)$ and touching the $k$ axis
first, while
the second one represents the average duration of a
flip (or success), the walker starting from $(0,N)$ and touching the $j$ axis
first.
Combining eqs.~(4.18, 4.19) and noting that
$$
\Tb={1\over b}+u_{1,N-1}
=v_{1,N-1}\, \left({1\over b}+u^{(k)}_{1,N-1}\right)
+(1-v_{1,N-1})\,\left({1\over b}+u^{(j)}_{1,N-1}\right)
\eqno(4.20)
$$
again yields the expression (4.15) of $\Tt$.

\vskip 14 pt
\noindent {\bf 5 Two limiting cases}
\vskip 12 pt

In this section we consider two simple limiting cases
$a=0$ and $b=0$, which allow for exact closed-form solutions,
that will be helpful to understand the general case.
We use here the discrete-time language, for simplicity,
since it yields the same equations and results
as the continuous-time formalism used so far, because
the hopping rates of the walker add to 1.

\bigskip
\noindent {\it 5.1 The case $a=0$}

In this limiting case the toy model reduces
to the classical {\it Gambler's ruin} problem on the segment $j+k=N$,
with equal probabilities $b=1/2$ for the walker
to hop to the right or to the left,
and with reflections on sites $(0,N)$ and $(N,0)$.
The stationary probability measure is uniform on that segment, i.e.,
$$
p_{j,k}={1\over N+1}\quad(j+k=N).
\eqno(5.1)
$$

On the other hand, eqs.~(4.10a) and (4.10b) have the following simple solutions
$$
u_{j,k}=jk,\quad v_{j,k}=\frac{k}{j+k}.
\eqno(5.2)
$$
By inserting these results into eqs.~(4.13) and (4.14), we are left with
$$
\Tb=N+1, \quad \Tt=D_{0,N}=N(N+1).
\eqno(5.3)
$$
The asymptotic $N^2$ law is due to the diffusive nature of the process.
One also finds
$$
D_{j,k}=N(N+1)-j(j+1).
\eqno(5.4)
$$

\medskip
\noindent {\it 5.2 The case $b=0$}

This case also yields a classical problem,
{\it Le probl\`eme des points}, considered by Fermat and Pascal
[14].

In this case, once the random walker has reached either of the axes,
it stays there.
Therefore for $b=0$ the model is not ergodic even for a finite system size,
and any probability
measure supported on the two points $(0,N)$ and $(N,0)$,
i.e., satisfying $p_{0,N}+p_{N,0}=1$, is stationary.
In the $b\to 0$ limit of the toy model we obtain the symmetric measure, i.e.,
$p_{0,N}=p_{N,0}=1/2$.

Let us now compute $\u$ and $\v$.
Consider the more general case
where the probability of hopping south is $p$ and of hopping west is $r$,
with $p+r=1$.
Eqs.~(4.10) and (4.11) can be solved by means of generating functions.
Dealing first with the $v_{j,k}$,
we introduce the two-variable generating series
$$
V(x,y)=\sum_{j=1}^\infty \sum_{k=1}^\infty v_{j,k}\ x^j\ y^k
.\eqno(5.5)
$$
Eq.~(4.10b) yields a closed equation for the function $V(x,y)$,
which can be solved explicitly.
We thus obtain the rational form
$$
V(x,y)={rxy\over(1-y)(1-rx-py)}
.\eqno(5.6)
$$
By expanding this expression as a double series, we obtain
$$
v_{j,k}=\sum_{n=0}^{k-1}\pmatrix{j+n-1\cr n}p^n\ r^j
.\eqno(5.7)
$$
Note that the $n\,$th term in the sum represents the probability
that the walker, starting at $(j,k)$,
reaches the $k$ axis after having made $n<k$ steps south.
In particular one gets
$$
v_{1,k}=1-p^k,\quad v_{j,1}=r^j
.\eqno(5.8)
$$

In a similar way, eq.~(4.10a) yields the following rational expression
for the generating series of the $\u$
$$
U(x,y)={xy\over(1-x)(1-y)(1-rx-py)}
.\eqno(5.9)
$$
By expanding this result we get
$$
u_{j,k}=\sum_{m=0}^{j-1}\sum_{n=0}^{k-1}\pmatrix{m+n\cr m}r^mp^n
,\eqno(5.10)
$$
and especially
$$
u_{1,k}={1-p^k\over r},\quad u_{j,1}={1-r^j\over p}
.\eqno(5.11)
$$

$\Tb$ and $\Tt$ are divergent when $b\to 0$.
Indeed
by inserting the above results for $p=r=1/2$ into eqs.~(4.13), (4.14),
we obtain their asymptotic behavior
in the $b\to 0$ regime:
$$
\Tb\approx {1\over b},\quad
\Tt\approx{2^{N-1}\over b}\quad(b\to 0).
\eqno(5.12)
$$
The second result has the form of eq.~(1.4),
with a mass $\mu=\ln 2$.

\medskip
Coming back to the original two-species exclusion process, one notes that
the two cases considered here correspond to two extreme situations.

The case $a=0$ corresponds to $\alpha$ infinite.
In this case the stationary state is known [5].
No holes are present in the system and
the remaining  $(+)$ and $(-)$ particles play the role
of particles and holes, respectively, in the
equivalent one-species totally asymmetric exclusion process
to which the original process reduces.
Moreover, in this equivalent process, the rates
for input and output of particles
are equal to $\tilde \alpha=\beta/q$,
implying ---$\beta$ being small---that the system is at
coexistence between a low density phase with a density of particles
equal to $\tilde \alpha$ and a high density phase with a
density of particles equal to $1-\tilde \alpha$,
as reviewed in the Introduction.
The instantaneous configuration of the system consists of
two regions of order $N$, one of low density of $(+)$
(or high density of $(-)$)
on the left, separated from another one of high density of $(+)$
(or low density of $(-)$) on the right
by a {\it shock}.
The location of the shock diffuses between $0$ and $N$.
In the toy model,
the location of the random walker is precisely that of the shock.
It is uniformly distributed over the whole system.
As said in the Introduction this case does not correspond
to spontaneous symmetry breaking.

In the other extreme case $b=0$, i.e., $\alpha=0$,
the dynamics {\it stops} in one of the states of broken symmetry.
Hence this situation
does not correspond to genuine spontaneous symmetry breaking.

\vskip 14 pt
\noindent {\bf 6 The general case}
\vskip 12 pt

We now turn to the analysis of the large-$N$ behavior
of the stationary probabilities $p_{j,k}$
and of the average  time $\Tt$ between two flips
in the general situation $(0<a<1/2$, $b=1/2-a)$.

\bigskip
\noindent {\it 6.1 The stationary probability measure}

The stationary probability measure $p_{j,k}$ is the solution of eq.~(4.2),
with the boundary conditions~(4.3).
It is convenient to use as co-ordinates $j$ and the number of holes $\nu$,
such that $j+k+\nu=N$.
With the notation $p_j^{(\nu)}=p_{j,k}$, eq.~(4.2) reads
$$
p_j^{(\nu)}= a\left(p_j^{(\nu-1)}+p_{j+1}^{(\nu-1)}\right)
+b\left(p_{j+1}^{(\nu)}+p_{j-1}^{(\nu)}\right).
\eqno(6.1)
$$
with boundary conditions inherited from eq.~(4.3).
It is shown in Appendix B how this equation can be solved numerically
in a recursive way,
according to increasing values of $\nu$, starting from $\nu=0$.

In the asymptotic $N\to\infty$ regime of interest,
the probability measure is expected to be concentrated
in two finite regions around the points $(0,N)$ and $(N,0)$,
each region being decoupled from the other one.
This phenomenon, already observed in section 5 for $b=0$,
is a consequence of the bias present in the dynamical rules
of the toy model.

Because of the symmetry property (4.4),
it is sufficient to consider the vicinity of the point $(0,N)$.
We are thus led to look for an asymptotic solution $p_j^{(\nu)}$ of eq.~(6.1),
independent of $N$ in the $N\to\infty$ limit,
and localized, namely decreasing as either $j$ or $\nu$ gets large,
i.e., for $1\ll j,\nu\ll N$.
The normalization of the probability measure now reads
$$
\sum_{\nu\ge 0}\sum_{j\ge 0}p_j^{(\nu)}={1\over 2},
\eqno(6.2)
$$
since the two decoupled regions around $(0,N)$ and around $(N,0)$
bear equal weights.

In order to study these asymptotic probabilities $p_j^{(\nu)}$,
it is convenient to introduce the one-variable generating functions
$$
P_j(x)=\sum_{\nu\ge 0} p_j^{(\nu)}\, x^\nu.
\eqno(6.3)
$$
Eq.~(6.1) then leads to the recursion relation
$$
P_j(x)=ax\big[P_j(x)+P_{j+1}(x)\big]+b\big[P_{j+1}(x)+P_{j-1}(x)\big],
\eqno(6.4)
$$
with (see eq.~(4.3))
$$
P_0(x)=p_0^{(0)}.
\eqno(6.5)
$$
The general solution of eq.~(6.4) reads
$$
P_j(x)=A(x)[z_-(x)]^j+B(x)[z_+(x)]^j,
\eqno(6.6)
$$
where $z_\pm(x)$ are the two solutions of the characteristic equation
$$
(ax+b)z^2-(1-ax)z+b=0,
\eqno(6.7)
$$
hence
$$
z_{\pm}(x)=\frac{1-ax\pm\sqrt{\Delta(x)}}{2(ax+b)},
\eqno(6.8)
$$
where
$$
\Delta(x)=a^2x^2-4a(1-a)x+4a(1-a).
\eqno(6.9)
$$
This expression vanishes for $x=x_1$ and $x=x_2$, with
$$
x_1={2\over a}\left(1-a-\sqrt{(1-a)(1-2a)}\right),\quad
x_2={2\over a}\left(1-a+\sqrt{(1-a)(1-2a)}\right).
\eqno(6.10)
$$
The associated values of $z$ read
$$
\eqalign{
z_1=z_\pm(x_1)&={1-2a+2\sqrt{(1-a)(1-2a)}\over 3-2a},\cr
z_2=z_\pm(x_2)&={1-2a-2\sqrt{(1-a)(1-2a)}\over 3-2a}.
}
\eqno(6.11)
$$

The branch $z_-(x)$ is monotonically increasing
from $z_-=\exp(-\sigma)$ for $x=0$, with the notation
$$
\cosh\sigma={1\over 2b}={1\over 1-2a},\ \
{\rm i.e.,}\ \
\exp(\pm\sigma)={1\pm\sqrt{4a(1-a)}\over 1-2a}
\eqno(6.12)
$$
to $z_-=z_1$ for $x=x_1$.
Conversely, $z_+(x)$ is monotonically decreasing
from $z_+=\exp(\sigma)$ for $x=0$ to $z_+=z_1$ for $x=x_1$.

Since the generating functions $P_j(x)$ are analytic at least
in a neighborhood of the origin $x=0$,
and decrease to zero for large $j$, the second term in the general
solution~(6.6) is ruled out.
Taking into account the boundary condition~(6.5) at $j=0$,
we are left with
$$
P_j(x)=p^{(0)}_0[z_-(x)]^j.
\eqno(6.13)
$$
Then, setting $x=1$ in eq.~(6.13), and using $z_-(1)=1-2a$, yields
$$
P_j(1)=\sum_{\nu\ge 0} p_j^{(\nu)}=p^{(0)}_0(1-2a)^j.
\eqno(6.14)
$$
By summing this expression over $j$, and using eq.~(6.2),
one gets the very simple result
$$
p^{(0)}_0=a.
\eqno(6.15)
$$

The average time between two bounces $\Tb$ in the $N\to\infty$ regime
can then be evaluated by means of eq.~(4.16).
We thus obtain
$$
\Tb={1\over 2ab}.
\eqno(6.16)
$$

Eq.~(6.13) is the main result of this section,
from which the following consequences can be derived.

\medskip\noindent (i)
A quantitative characterization of the concentration of the probability measure
around the point $(0,N)$ (and similarly around $(N,0)$)
is provided by the following order parameter
$$
M_2(N)=\big\langle\left(\frac{j-k}{N}\right)^2\big\rangle,
\eqno(6.17)
$$
defined as an average with respect to the stationary measure.
The existence of the asymptotic probabilities $p_j^{(\nu)}$,
independent of $N$, implies
$$
M_2(N)=1-{\lambda\over N}+{\cal O}\left({1\over N^2}\right),
\quad\hbox{with}\;\lambda=4\langle j\rangle+2\langle\nu\rangle,
\eqno(6.18)
$$
the averages being taken with respect to the $p_j^{(\nu)}$.
The result~(6.13) yields
$$
\langle j\rangle=2\sum_{j\ge 0}j\,P_j(1)=\frac{1-2a}{2a}
\eqno(6.19)
$$
and
$$
\langle\nu\rangle=2\sum_{j\ge 0}\left(\frac{\d P_j(x)}{\d x}\right)_{x=1}
=\frac{(1-a)(1-2a)}{a}
\eqno(6.20)
$$
so that we have
$$
\lambda=\frac{2(1-2a)(2-a)}{a}.
\eqno(6.21)
$$
Note the simple interpretation of (6.19): $\langle j\rangle$, the
average number of $(-)$, i.e., of
minority particles in the steady state, is equal to the input rate $\alpha$.
It is also equal to ${\bf D}_{xy}/V_x$, using eqs.~(3.5, 3.6).

\medskip\noindent (ii)
{}From the expression~(6.13) of $P_j(x)$ it is also possible to extract
the $p_j^{(\nu)}$ by means of the following contour integral
encircling the origin in the complex $x$-plane
$$
p_j^{(\nu)}=\oint{\d x\over 2i\pi}{P_j(x)\over x^{\nu+1}}.
\eqno(6.22)
$$
We will restrict ourselves to the asymptotic behavior of these probabilities,
when both $j$ and $\nu$ are large $(1\ll j,\nu\ll N)$, at a fixed angle
$0\le\varphi\le\pi/4$ between the negative $k$-axis and the vector joining
$(0,N)$ to $(j,k)$, namely we set
$$
\left\{\eqalign{
N-k=j+\nu&=\rho\cos\varphi,\cr
j&=\rho\sin\varphi.\cr
}\right.
\eqno(6.23)
$$
We are thus led to evaluate the  integral
$$
p_j^{(\nu)}=a\oint{\d x\over 2i\pi x}\exp
\bigg\{-\rho\big[(\cos\varphi-\sin\varphi)\ln x-\sin\varphi\ln
z_-(x)\big]\bigg\},
\eqno(6.24)
$$
for $\varphi$ fixed, in the $\rho\to\infty$ limit.
The direction $\varphi=0$ corresponds to $\nu\to\infty$ at constant $j$,
whereas $\varphi=\pi/4$ corresponds to $j\to\infty$ at constant $\nu$.

A saddle-point calculation leads to the  estimate
$$
p_j^{(\nu)}\sim\exp\left\{-\rho\mu(\varphi)\right\},
\eqno(6.25)
$$
with
$$
\mu(\varphi)=(\cos\varphi-\sin\varphi)\ln x_c-\sin\varphi\ln z_-(x_c),
\eqno(6.26)
$$
and the saddle point $x_c(\varphi)$ is given by
$$
\cot\varphi=1+\frac{x_c}{z_-(x_c)}\left(\frac{\d z_-(x)}{\d x}\right)_{x=x_c}.
\eqno(6.27)
$$
We have thus demonstrated that the asymptotic probabilities $p_j^{(\nu)}$
decay exponentially in all directions away from the point $(0,N)$,
with an inverse decay length given by the {\it mass}
$\mu(\varphi)$ in the direction defined by the angle $\varphi$.
Eqs.~(6.26) and (6.27) provide a parametric representation of $\mu(\varphi)$.
It increases monotonically between $\varphi=0$,
where $x_c=x_1$, and
$$
\mu(0)=\mu=\ln x_1,
\eqno(6.28)
$$
with the notation~(6.10),
and $\varphi=\pi/4$, where $x_c=0$, and $\mu(\pi/4)=\sigma/\sqrt{2}$,
with the notation~(6.12).

\bigskip
\noindent {\it 6.2 The average time between two flips $\Tt$}

This quantity is given by eq.~(4.13).
First,
the asymptotic behavior of $u_{1,N-1}$ for large $N$ is easy to determine.
Indeed the walker will visit the $j$-axis with a vanishingly small probability.
One may therefore neglect the presence of this boundary,
and just consider the displacements of the walker parallel to the $j$-axis.
Thus $u_{j,k}$ reduces to $u_j$, the average time for the walker starting
at site $j$ to reach the boundary, i.e., site $j=0$.
During $\d\tau$ the walker may either hop to the left with
probability $(a+b)\d\tau$ or to the right with probability $b\d\tau$
or stay on site $j$ with probability $a\d\tau$.
Therefore the set of $u_j$ obeys the  equation
$$
u_j=1+(a+b)u_{j-1}+bu_{j+1}+au_j
,\eqno(6.29)
$$
with the boundary condition $u_0=0$.
Equivalently eq.~(6.29) is obtained by dropping the $k$-dependency
in eq.~(4.10a).
The solution to eq.~(6.29) is simply $u_j=j/a$.
Coming back to the original two-dimensional situation, we find
$$
u_{1,N-1}\to {1\over a}\quad(N\gg 1).
\eqno(6.30)
$$
This result provides an alternative derivation of eq.~(6.15),
by means of eq.~(4.17).

The analysis of the asymptotic behavior of $v_{j,1}$ is more difficult.
Introducing the generating function $V(x,y)$ as in eq.~(5.5),
we see that eq.~(4.10b) is equivalent to
$$
D(x,y)V(x,y)+bxy\big[xV_1(x)+yV_2(y)\big]={x^2y^2\over 2(1-y)}
,\eqno(6.31)
$$
where
$$
\left\{\eqalign{
V_1(x)&=\p{V}{y}(x,y)_{y=0}=\sum_{j=1}^\infty v_{j,1}\, x^j ,\cr
V_2(y)&=\p{V}{x}(x,y)_{x=0}=\sum_{k=1}^\infty v_{1,k}\, y^k ,\cr
}\right.
\eqno(6.32)
$$
and
$$
D(x,y)=xy-a(x+y)xy-b(x^2+y^2)
.\eqno(6.33)
$$

The characteristic equation $D(x,y)=0$
defines the characteristic curve of the problem.
It is a rational cubic curve, with the origin as a double point,
and the three lines $x=-b/a$, $y=-b/a$, and $x+y=(1+2b)/a$ as asymptotes.
Figure 1 depicts the characteristic curve for $a=0.4$, $b=0.5-a=0.1$.

We notice that the characteristic equation is equivalent to eq.~(6.7),
up to the substitution $y=xz$.
Solving the characteristic equation for $y$, we thus get
$$
y_\pm(x)=xz_\pm(x),
\eqno(6.34)
$$
with the notation~(6.8).

The characteristic curve has
either a horizontal or a vertical tangent
at the four points marked on Figure 1.
Their co-ordinates read
$A=(x_1,y_1)$, $B=(y_1,x_1)$, $C=(x_2,y_2)$, $D=(y_2,x_2)$,
where $y_1=x_1z_1$, $y_2=x_2z_2$,
with the notations~(6.10) and (6.11).
Figure 1 also provides a geometric illustration of the saddle-point approach
to the stationary probability, presented in section 6.1.
Indeed, as the angle $\varphi$ varies from 0 to $\pi/4$,
the saddle point runs along the dispersion curve,
 in co-ordinates
$\big(x_c, y_c\big)=\big(x_c, x_c z_-(x_c)\big)$,
from the point $A$ to the origin $O$.

Let us come back to the $v_{j,k}$.
Since these numbers are bounded,
the function $V(x,y)$ is analytic in a complex domain ${\cal D}$
containing at least the product of disks $\vert x\vert<1$, $\vert y\vert<1$.
Whenever $x$ and $y$ are in ${\cal D}$, and related
by the characteristic equation $D(x,y)=0$, eq.~(6.31) simplifies to
$$
xV_1(x)+yV_2(y)={xy\over 2b(1-y)}
.\eqno(6.35)
$$
This equation shows that at least
one of the two functions $V_1(x)$ or $V_2(y)$ becomes singular
when the relation between $x$ and $y$ becomes itself singular,
i.e., at the marked points mentioned above, as may be seen
by taking the derivatives of both sides.

The points $A$ and $B$ are the relevant ones,
since they are the closest to the origin.
Since these two points play symmetric roles,
we can consider point $A$ for definiteness.
We have $y_1<1<x_1$, so that $V_1(x)$ is singular at $x=x_1$,
whereas $V_2(y)$ is regular at $y=y_1$.

We set
$$
x=x_1-\epsilon_x,\quad y=y_1+\epsilon_y
,\eqno(6.36)
$$
so that the characteristic equation reads, to leading order in
$\epsilon_x$ and $\epsilon_y$
$$
\epsilon_x\approx\left({ax_1+b\over x_1}\right)^2
{\epsilon_y^2\over aw},
\quad
{\rm with}\quad
w=\sqrt{(1-a)(1-2a)}
.\eqno(6.37)
$$
By inserting this estimate into eq.~(6.35),
and expanding both sides to leading order in $\epsilon_y$,
we are led to the conclusion that $V_1(x)$ has a singular part of the form
$$
V_1(x)\approx V_1(x_1)-K\sqrt{x_1-x}
,\eqno(6.38)
$$
with
$$
K={\sqrt{aw}\over ax_1+b}\p{}{y}
\left[yV_2(y)-{x_1y\over 2b(1-y)}\right]_{y=y_1}
.\eqno(6.39)
$$

By inserting the singular behavior~(6.38) into the definition~(6.32)
of $V_1(x)$, we get  the  estimate
$$
v_{j,1}\approx{K\exp(\mu /2)\over 2\sqrt{\pi}}
j^{-3/2}\exp(-j\mu)\quad(j\gg 1)
,\eqno(6.40)
$$
where $\mu$ has been defined in eq.~(6.28).

We can now insert both estimates~(6.30) and (6.40)
into the expression~(4.13) of $\Tt$.
We thus obtain
$$
\Tt\approx L\, N^{3/2}\exp(N\mu)
\quad(N\gg 1)
,\eqno(6.41)
$$
with
$$
L={2\sqrt{\pi}\over a(1-2a)K
\exp(3\mu /2)}
.\eqno(6.42)
$$
We have thus derived the full asymptotic expression
of the average time between two flips.

\medskip
\noindent $\bullet$ The mass $\mu$ is given by eqs.~(6.10), (6.28).
It vanishes for small values of $a$ as $\mu=a/4+11a^2/32+\cdots$
This regime corresponds to a continuum limit,
to be investigated in detail in section 7.
The opposite limit $a\to 1/2$ yields $\mu\to\ln 2$, in agreement with
eq.~(5.12).
The mass increases monotonically between these two limits.
For instance in the case $a=b=1/4$
one finds $\mu=\ln\big(2(3-\sqrt{6})\big)=0.096\, 237$.

\medskip
\noindent $\bullet$ The exponent 3/2 is universal,
i.e. independent of $a$,
except for the two limiting cases $a=0$ and $b=0$,
investigated in section 5.
The general analysis presented above fails in these cases,
because the characteristic curve is degenerate.
We have indeed $D(x,y)=-(x-y)^2/2$ for $a=0$,
and $D(x,y)=xy(2-x-y)/2$ for $b=0$.

\medskip
\noindent $\bullet$ The absolute prefactor $L$ only depends on $a$.
It cannot be determined within the present approach.
The analysis of the continuum limit
will provide a quantitative estimate of the behavior
of this prefactor as $a\to 0$.
Indeed eq.~(7.45) yields $L\approx2/\sqrt{\pi^3 a}$,
i.e., $K\approx\pi^2/\sqrt{a}$.

\vskip 14 pt
\noindent {\bf 7 The continuum limit}
\vskip 12 pt

In this section we determine the scaling behavior of
$\u$, $\v$ and $\Tt$
in the continuum limit, corresponding to the regime of a small bias
$(a\to 0)$.
It will turn out that this analysis provides a complete description
of the crossover between the diffusive power law~(5.3)
and the exponential law~(6.41).

We consider first the probability of ruin $\v$.
We assume that $\v$ is a smooth function
of the continuous rescaled variables $X=ja$ and $Y=ka$.
Expanding eq.~(4.10b) to the lowest non-trivial order in $a$ leads to
$$
\left({\partial\over\partial X}-
{\partial\over\partial Y}\right )^2 v(X,Y)
=2\left({\partial\over\partial X}+
{\partial\over\partial Y}\right ) v(X,Y)
,\eqno(7.1)
$$
hence to
$$
\pp {v(\xi,\theta)}\xi=\p {v(\xi,\theta)}\theta
,\eqno(7.2)
$$
with
$$
\left\{\eqalign{
&\xi=X-Y=(j-k)a,\cr
&\theta=X+Y=(j+k)a.
}\right.
\eqno(7.3)
$$
We thus obtain a heat equation, or diffusion equation,
where $\xi$ and $\theta$ are respectively the space-like and the time-like
variables.
Note that the time-like axis is oriented along the direction of the bias.
This partial differential equation inherits its
boundary conditions from eqs.~(4.11a, b)
$$
v(\xi=-\theta,\theta)=1,\quad v(\xi=\theta,\theta)=0
.\eqno(7.4)
$$
These moving boundary conditions are difficult to handle.
It is convenient to introduce the dimensionless space-like variable
$$
x={k\over j+k}
,\eqno(7.5)
$$
such that
$$
\xi=(1-2x)\theta
.\eqno(7.6)
$$
Eq.~(7.2) thus transforms into
$$
\pp{v}{x}-2(1-2x)\theta\p{v}{x}-4\theta^2\p{v}{\theta}=0
,\eqno(7.7)
$$
with the boundary conditions
$$
v(x=0,\theta)=0,\quad v(x=1,\theta)=1
.\eqno(7.8)
$$

The Gaussian
$$
G(x,\theta)=(4\pi\theta)^{-1/2}\exp\big(-\theta(x-1/2)^2\big)
\eqno(7.9)
$$
is an elementary solution of eq.~(7.7), neglecting boundary conditions.
We are therefore led to the change of function
$$
v(x,\theta)=G(x,\theta)F_v(x,\theta)
,\eqno(7.10)
$$
which inserted into eq.~(7.7) gives
$$
\pp{F_v}{x}-4\theta^2\p{F_v}{\theta}=0
.\eqno(7.11)
$$
A last change on the time-like variable
$$
\theta=-{1\over 4\eps}
\eqno(7.12)
$$
leads to a heat equation for $F_v$,
$$
\pp{F_v}{x}=\p{F_v}{\eps}
,\eqno(7.13)
$$
with boundary conditions
$$
F_v(x=0,\eps)=0,\quad
F_v(x=1,\eps)=f_v(\eps)=\left({\pi\over-\eps}\right)^{1/2}
\exp\left(-{1\over 16\eps}\right)
.\eqno(7.14)
$$
and with $\lim_{\eps\to-\infty}F_v(x,\eps)=0$.

We thus have to solve the heat equation on the unit interval,
with prescribed time-dependent boundary conditions.
The solution is given by the convolution
$$
F_v(x,\eps)=\int_0^\infty\d\eps'\,f_v(\eps-\eps')\,g_v(x,\eps')
,\eqno(7.15)
$$
where
$$
g_v(x,\eps)=
2\pi\sum_{k=1}^\infty(-1)^{k-1}\,k\,\sin(k\pi x)\,\exp(-k^2\pi^2\eps)
\eqno(7.16)
$$
is the appropriate Green's function of the heat equation (7.13).
We mention for further reference its alternative expression,
obtained from eq.~(7.16) by the Poisson formula
$$
g_v(x,\eps)=(4\pi\eps^3)^{-1/2}
\sum_{m=-\infty}^{\infty}(2m+1-x)\exp\left(-{(2m+1-x)^2\over 4\eps}\right)
.\eqno(7.17)
$$

We are thus left with the following explicit formula
for the scaling behavior of the $\v$ in the continuum limit
$$
\v\approx v(x,\theta)=\exp\big(-\theta(x-1/2)^2\big)
\int_0^{\infty}\, {\d\eps\over(1+4\theta\eps)^{1/2}}
\exp\left({\theta\over 4(1+4\theta\eps)}\right)\,g_v(x,\eps)
.\eqno(7.18)
$$
We recall that the scaling variables $x$ and $\theta$
are related to $j$ and $k$ by eqs.~(7.3, 7.5).

We are especially interested in the quantity $v_{N-1,1}$,
which enters the expression (4.13) of $\Tt$.
Eq.~(7.18) yields
$$
v_{N-1,1}\approx a\Phi(\theta)\quad(\theta=Na)
,\eqno(7.19)
$$
where $\Phi(\theta)$ is the following scaling function
$$
\Phi(\theta)={1\over\theta}\p{}{x}v(x=0,\theta)
={1\over\theta}\int_0^{\infty}\, {\d\eps\over(1+4\theta\eps)^{1/2}}
\exp\left(-{\theta^2\eps\over 1+4\theta\eps}\right)\,g_0(\eps)
,\eqno(7.20)
$$
with
$$
\eqalign{
g_0(\eps)&=\p{}{x}g_v(x=0,\eps)\cr
&=2\pi^2\sum_{k=1}^\infty(-1)^{k-1}\,k^2\,\exp(-k^2\pi^2\eps)\cr
&=(4\pi\eps^3)^{-1/2}
\sum_{m=-\infty}^{\infty}\left( {(2m+1)^2\over 2\eps}-1\right)
\exp\left(-{(2m+1)^2\over 4\eps}\right).\cr
}\eqno(7.21)
$$

The scaling behavior of the duration of the game
$\u$ can be investigated in a similar way.
Expanding eq.~(4.10a) now leads to an inhomogeneous heat equation,
$$
\p {u(\xi,\theta)}\theta-\pp {u(\xi,\theta)}\xi ={1\over 2a^2}
,\eqno(7.22)
$$
with the moving boundary conditions
$$
u(\xi=-\theta,\theta)=u(\xi=\theta,\theta)=0
.\eqno(7.23)
$$
We again use the dimensionless space-like variable $x$, setting
$$
u(x,\theta)=G(x,\theta)F_u(x,\theta)
,\eqno(7.24)
$$
and we change the time-like variable from $\theta$ to $\eps$.
We thus obtain an inhomogeneous heat equation for $F_u$,
$$
\p{F_u}{\eps}-\pp{F_u}{x}=f_u(x,\eps)
,\eqno(7.25)
$$
with boundary conditions
$$
F_u(x=0,\eps)=F_u(x=1,\eps)=0
,\eqno(7.26)
$$
and with $\lim_{\eps\to-\infty}F_u(x,\eps)=0$.
The source term reads
$$
f_u(x,\eps)={1\over 8a^2}\left({\pi\over(-\eps)^5}\right)^{1/2}
\exp\left(-{(2x-1)^2\over 16\eps}\right)
.\eqno(7.27)
$$

We now have to solve the heat equation on an interval,
with a prescribed time-dependent source, and Dirichlet boundary conditions.
The solution is again given by a convolution,
$$
F_u(x,\eps)=\int_0^\infty\d\eps'\int_0^1\d y\,f_u(y,\eps-\eps')\,g_u(x,y,\eps')
,\eqno(7.28)
$$
where
$$
g_u(x,y,\eps)=
2\sum_{k=1}^\infty\sin(k\pi x)\,\sin(k\pi y)\exp(-k^2\pi^2\eps)
\eqno(7.29)
$$
is the appropriate Green's function of the heat equation.

We are thus left with the following explicit formula
for the scaling behavior of the $\u$ in the continuum limit
$$
\eqalign{
\u\approx u(x,\theta)&={2\theta^2\over a^2}\exp\big(-\theta(x-1/2)^2\big)\cr
&\times\int_0^1\d y\int_0^{\infty}\, {\d\eps\over(1+4\theta\eps)^{5/2}}
\exp\left({\theta(2y-1)^2\over 4(1+4\theta\eps)}\right)\,g_u(x,y,\eps).
}
\eqno(7.30)
$$

We are especially interested in the quantity $u_{N-1,1}$,
which enters the expression (4.13) of $\Tt$.
Eq.~(7.30) yields
$$
u_{N-1,1}\approx {1\over a}\Psi(\theta)\quad(\theta=Na)
,\eqno(7.31)
$$
where $\Psi(\theta)$ is the following scaling function
$$
\Psi(\theta)={a^2\over\theta}\p{}{x}u(x=0,\theta)
=2\theta\int_0^1\d y\int_0^{\infty}\, {\d\eps\over(1+4\theta\eps)^{5/2}}
\exp\left({(y^2-y-\theta\eps)\theta\over 1+4\theta\eps}\right)\,g_v(1-y,\eps)
,\eqno(7.32)
$$
where we have used the identity
$$
\p{}{x}g_u(x=0,y,\eps)=g_v(1-y,\eps)
.\eqno(7.33)
$$

Putting together both results~(7.19, 7.31)
we obtain the following expression for the scaling behavior
of the average time between two flips
$$
\Tt\approx{1\over a^2}\Delta(\theta)\quad(\theta=Na)
,\eqno(7.34)
$$
where the scaling function $\Delta(\theta)$ reads
$$
\Delta(\theta)={\Psi(\theta)\over\Phi(\theta)}
.\eqno(7.35)
$$

The regimes of small and large values of the time-like scaling
variable $\theta=Na$ are of special interest.
We now show how the scaling law~(7.34)
describes the crossover between the diffusive law~(5.3) in the unbiased case
$(\theta\ll 1)$,
and the exponential law~(6.41) in the generic biased case
$(\theta\gg 1)$.

\noindent $\bullet$ $\theta\ll 1$.

In this regime it is most convenient to consider eqs.~(7.2, 7.22)
and to look for solutions to them in the form of power series in $\theta$.
We thus obtain
$$
\left\{\eqalign{
u(x,\theta)&={\theta^2\over a^2}x(1-x)
\left(1-\theta+{\theta^2\over 3}(4+x-x^2)+\cdots\right),\cr
v(x,\theta)&=x+x(1-x)(1-2x)
\left(-{\theta\over 3}+{\theta^2\over 15}(1-2x+2x^2)+\cdots\right).\cr
}\right.\eqno(7.36)
$$
The leading terms of the above expressions agree with the results~(5.2)
corresponding to $a=0$.
The above results also yield the following expansions
for the scaling functions
$$
\left\{\eqalign{
\Phi(\theta)&={1\over\theta}\left(1-{\theta\over 3}+{\theta^2\over
15}+\cdots\right),\cr
\Psi(\theta)&=\theta\left(1-\theta+{4\theta^2\over 3}+\cdots\right),\cr
\Delta(\theta)&=\theta^2\left(1-{2\theta\over 3}+{47\theta^2\over
45}+\cdots\right).\cr
}\right.\eqno(7.37)
$$
The last of these expansions yields more explicitly
$$
\Tt\approx N^2\left(1-{2Na\over 3}+{47N^2a^2\over 45}+\cdots\right)
\quad(N\gg 1, Na\ll 1)
.\eqno(7.38)
$$
The leading term reproduces the asymptotic behavior of the exact result~(5.3),
characteristic of diffusive motion.
Note that the first correction term to this law is negative.
This may be understood as follows.
In the current regime
(i.e., $\alpha\gg N\gg 1$),
we typically expect a small number of zeros to separate the
$(+)$ and $(-)$ blocks, thus leading to a small negative correction
to the average time between two flips of the strictly diffusive shock behavior.

\noindent $\bullet$ $\theta\gg 1$.

This regime is the most relevant physically,
since it corresponds to the asymptotic behavior~(6.41).
The asymptotic behavior of the scaling functions
can be determined from the integral expressions~(7.20, 7.32).

The calculation for $\Psi(\theta)$ is as follows.
The double integral in eq.~(7.32) is dominated
by the region where both variables $\epsilon$ and $y$ are small.
In this regime the function $g_v(1-y,\eps)$ can be approximated
by the $m=0$ term in the sum~(7.17).
Choosing the rescaled integration variables
$z=y/(\theta\eps)$ and $\zeta=\theta^2\eps$, we are left with
$$
\Psi(\theta)\approx{1\over\sqrt{\pi}}\int_0^\infty z\d z
\int_0^\infty{\zeta^{1/2}\d\zeta\over(1+4\zeta/\theta)^{5/2}}
\exp\left(-{(z+2)^2\zeta\over 4(1+4\zeta/\theta)}\right)
.\eqno(7.39)
$$
The $\zeta$-integral can be performed by changing variable
from $\zeta$ to $\eta=\zeta/(1+4\zeta/\theta)$.
After some algebra we are left with the very simple limit behavior
$$
\Psi(\theta)\to 1\quad(\theta\gg 1)
,\eqno(7.40)
$$
in agreement with the more general result (6.30).
The correction to this limit vanishes exponentially as $\exp(-\theta/4)$,
since it is due to the upper bound $\eta_{\rm max}=\theta/4$
of the $\eta$-integral.

Let us now compute $\Phi(\theta)$.
We perform the change of variable $\zeta=\theta/(1+4\theta\eps)$
in the integral representation~(7.20), with the last expression
of eq.~(7.21)
for the function $g_0$, obtaining thus
$$
\Phi(\theta)\approx{2\exp(-\theta/4)\over(\pi\theta^3)^{1/2}}I
,\eqno(7.41)
$$
with
$$
\eqalign{
I&=\sum_{m=0}^{\infty}
\int_0^\infty\d\zeta\big(2(2m+1)^2\zeta-1\big)
\exp\big[-\big((2m+1)^2-1/4\big)\zeta\big]\cr
&=\sum_{m=0}^\infty{(2m+1)^2+1/4\over\big((2m+1)^2-1/4\big)^2}
=2\sum_{m=0}^\infty\left({1\over(4m+1)^2}+{1\over(4m+3)^2}\right)
={\pi^2\over4}.\cr
}\eqno(7.42)
$$
We thus obtain the estimate
$$
\Phi(\theta)
\approx{1\over
2}\left({\pi\over\theta}\right)^{3/2}
\exp(-\theta/4)\quad(\theta\gg 1)
.\eqno(7.43)
$$

The asymptotic estimates~(7.40) and (7.43) yield
$$
\Delta(\theta)\approx
2\left({\theta\over\pi}\right)^{3/2}
\exp(\theta/4)\quad(\theta\gg 1)
,\eqno(7.44)
$$
namely
$$
\Tt\approx 2\left({N^3\over a\pi^3}\right)^{1/2}\exp(Na/4)
\quad(Na\gg 1)
.\eqno(7.45)
$$
This result exhibits an exponential growth
with mass $\mu=a/4$ and
a power-law prefactor in $N^{3/2}$.
These results are in agreement with the analysis of section 6.

In conclusion we see that the analysis contained in this section fully
describes
the crossover from the diffusive behavior of a shock
($\theta\ll 1$, i.e., $\alpha\gg N\gg 1$)
to the symmetry breaking regime ($\theta\gg 1$, i.e., $N\gg\alpha\gg 1$).
This crossover behavior is illustrated on Figure 2,
showing a logarithmic plot of $\Tt$ against $\theta=Na$.
The numerical data for $a=0.1$, 0.06, and 0.03,
obtained by means of the approach described in Appendix B,
smoothly converge toward the result~(7.35) of the continuum limit.
The smoothness of the corrections to the continuum limit
has already been noticed below eq.~(6.42), for the mass $\mu$.

It is striking to observe the richness of the behavior of the
original exclusion process: even in the regime studied here,
i.e., $\beta \to 0$ and $\alpha\to \infty$, the curve obtained
for $\Tt$ is very similar to that obtained in [4]
by numerical simulations of the exclusion process with finite values
of $\alpha$ and $\beta$ (see Fig.~4 of that work).

Finally we notice that
the stationary probability measure $p_{j,k}$ also exhibits
a scaling behavior in the continuum limit.
A detailed investigation of this behavior is not needed.
We just mention that eq.~(4.17) implies the following
scaling law for the probability at both endpoints $(0,N)$ and $(N,0)$,
in the regime $a\ll 1$, $N\gg 1$
$$
p_{0,N}=p_0^{(0)}\approx{a\over\Psi(\theta)}\quad(\theta=Na).
\eqno(7.46)
$$
As expected, this result interpolates between the $a=0$ limit,
 eq.~(5.1), and the $N\to\infty$ regime,  eq.~(6.15).

\vskip 14 pt
\noindent {\bf 8 Conclusion}
\vskip 12 pt
\noindent

The analysis presented in this paper confirms the intuitive picture,
given at the end of section 3,
of the mechanism leading to spontaneous symmetry breaking in the toy model,
and quantifies it in several respects.

First, the stationary probability measure $p_{j,k}$
is localized in a symmetric way
around the endpoints $(0,N)$ and $(N,0)$,
in the asymptotic regime of a large system size $N$.
This is intuitively clear:
the bias of the random walker is in the south-west direction
so that its drift is towards whichever boundary is nearest.
It then becomes localized near one of the endpoints due
to the bouncing property of the boundary.
This localization is quantified by the behavior~(6.18)
of the order parameter $M_2(N)$,
and by the exponential decay~(6.25) of the stationary probabilities
in all directions away from both endpoints.
The mass, or inverse decay length, $\mu(\varphi)$ depends on
the orientation:
it decreases monotonically
between $\varphi=\pi/4$ (decay along the diagonal $j+k=N$)
and $\varphi=0$ (decay along both co-ordinate axes).

This point brings some more insight on the flipping mechanism.
Since the decay length is the largest along the axes,
the walker, starting from $(0,N)$,
will preferentially diffuse down the $k$-axis,
until it possibly reaches the vicinity of the origin,
corresponding in the toy model
to few $(-)$, few $(+)$ and many holes in the system.
Then, in a few steps, it can pass the line $j=k$ and reach a region
where the bias drives it to the $j$-axis.
The typical duration of such a successful path, from $(0,N)$
down to the $j$-axis, is presumably of order $N$, indicating
that, when the system flips, it does so on a very short time scale
compared to the average duration of time it spends in a long lived state,
i.e., near the two endpoints $(0,N)$ or $(N,0)$.

This argument also provides an estimate of the average time
between two flips $\Tt$, i.e., the average lifetime of the walker in one of
the two regions near the endpoints.
Indeed the probability that the walker reaches the vicinity of the origin
is proportional to $\exp(-N\mu)$,
with $\mu=\mu(0)$,
therefore $\Tt$ should be
proportional to $\exp(N\mu)$.

We indeed find that the
average time $\Tt$ between two flips in the toy model reads
$$
\Tt\approx L\,N^{3/2}\,\exp(N\mu),
\eqno(8.1)
$$
with
$$
\mu=\mu(0)=\ln x_1=\ln\left[
{2\over a}\left(1-a-\sqrt{(1-a)(1-2a)}\right)
\right].
\eqno(8.2)
$$
This result agrees with the estimate~(1.4).
It also confirms the above picture of the flipping mechanism.
The mass $\mu$ can be interpreted as an activation energy,
or barrier height per particle;
the exponent $3/2$ of the power-law prefactor is universal,
i.e., it does not depend on the strength of the bias;
the absolute prefactor $L$ is only known
in the continuum limit of a weak bias ($a\to 0$),
where $L\approx 2/\sqrt{a\pi^3}$.

We have thus demonstrated the occurrence of spontaneous symmetry breaking in
the $\beta\to 0$ limit of the two-species exclusion model,
in the sense that the system flips between two symmetry-related
long lived states,
by showing that the two properties mentioned above hold.

\bigskip
Let us now impose boundary conditions which favor one species
of particles over the other,
breaking thus the symmetry explicitly.
We consider an exclusion model where, instead of having boundary rates
$\alpha, \beta$ for both types of particles,
we have $\alpha, \beta_+$ for the positive particles
and $\alpha, \beta_-$ for the negative particles.
We set $\beta_\pm=\beta(1\mp H)$,
where $H$ is a symmetry-breaking field, such that $\vert H\vert<1$.
Hence the positive (respectively, negative) particles
are favored by the boundary conditions
for $H>0$ (respectively, $H<0$).
If we now take $\beta\to 0$,
it can easily be checked that we find a toy model
with the following hopping rates: $a(1+H)$ west; $a(1-H)$ south;
$b(1+H)$ north-west; $b(1-H)$ south-east, again with $a+b=1/2$.

The bias associated with this model now reads
$$
{\bf V}=a\pmatrix{-1\cr -1}+(1-a)H\pmatrix{-1\cr 1}.
\eqno(8.3)
$$
As long as the field is small enough, i.e., $\vert H\vert <H_0=a/(1-a)$,
both components of the bias are still negative,
and the mechanism which localizes the walker near the endpoints is retained.
The significance of this is as follows.
Assume now $H>0$ for definiteness.
Even though the boundary conditions favor blocks of positive particles,
the system may still spend lengths of time
which diverge exponentially with the system size
in a state dominated by negative particles.
It can indeed be shown that the average durations of
time spent in the $(+)$ phase
and in the $(-)$ phase scale respectively as
$\Tt_+\sim\exp(N\mu_+)$ and $\Tt_-\sim\exp(N\mu_-)$,
where $\mu_+>\mu_-$ are the masses along both axes, which are now different.
This reinforces our proposed interpretation of the mass
as the activation energy per particle,
in the sense that $\Delta\mu=\mu_+-\mu_-$ is naturally interpreted
as the difference in energy per particle between both pure phases.
In the continuum limit of the model,
defined as $a\to 0$ and $H\to 0$ simultaneously,
we have $\mu_\pm\approx(a\pm H)^2/(4a)$, hence $\Delta\mu\approx H$.

The analogy between $H$ and a symmetry-breaking field in equilibrium
statistical physics, such as a magnetic field in the Ising model,
can be pursued.
Consider the order parameter $M_1(N)=\langle(k-j)/N\rangle$,
analogous to a magnetization.
This quantity can be estimated by noticing that the statistical weights
$p_\pm$ of both phases are proportional to the durations of time
$\Tt_\pm$ spent in them,
hence $M_1(N)\approx p_+-p_-\approx\tanh(NH/2)$ within exponential
accuracy,
the latter formula being identical to the equation of state
of the Ising model near coexistence.

We finally notice that the unfavored phase is metastable for $H<H_0$
and just unstable for $H>H_0$.
In the latter case the system will always mostly consist of $(+)$ particles.

We close up with a short discussion on the behavior of the mass
in the continuum limit of a weak bias.
We consider the most general biased random walk in the quarter plane,
where the velocity is still given by eq.~(3.5),
while the diffusion coefficients $\Dpar$ and $\Dperp$ are now
different from each other a priori.
It can be shown that the on-axis mass $\mu$ which enters the law~(1.4)
vanishes linearly with $a$,  with a non universal prefactor
depending on the diffusion coefficients,
$$
\mu\approx{a\over 2\Dperp\left(1+\sqrt{1+\Dpar/\Dperp}\,\right)}.
\eqno(8.4)
$$
In the case of the toy model, with $\Dpar\approx 0$ and $\Dperp\approx 1$,
we recover $\mu\approx a/4$ in the continuum limit.
Another case of interest is the north-south-east-west model,
defined by moves to the four cardinal directions
with rates $b$ (north and west), $a+b$ (south and east),
(with $a+2 b=1/2$),
for which interesting properties can be derived by means
of several approaches [15, 16].
In that case, we have $\Dpar\approx\Dperp\approx 1/4$,
hence $\mu\approx 2(\sqrt2-1)a$ in the continuum limit.

\bigskip
\noindent {\bf Acknowledgements}
\smallskip
We wish to thank F. Baccelli, G. Fayolle, J.L. Lebowitz, K. Mallick, and J.
Neveu for interesting conversations.
MRE, CG, DM and ERS thank the Isaac Newton Institute, Cambridge, for
hospitality.
DM and SS acknowledge support of Minerva Foundation, Munich.

\vfill\break
\noindent{\bf References}
\vskip 12pt
{\parindent 0em

[1] See for example Parisi G., {\it Statistical Field Theory},
Addison-Wesley (1988),
and references therein.

[2] See for example van Leeuwen J. M. J. and Hilhorst H. J.,
Physica {\bf A 107}, 319 (1981).
A review is given in: Forgacs G., Lipowsky R., and Nieuwenhuizen Th. M.,
in {\it Phase Transitions} (Domb C. and Lebowitz J. L., eds.),
vol. {\bf 14}, Academic Press (1991).

[3] Gacs P., J. Comput. Sys. Sci. {\bf 32}, 15 (1986).

[4] Evans M. R., Foster D. P., Godr\`eche C., and Mukamel D.,
Phys. Rev. Lett. {\bf 74}, 208 (1995).

[5] Evans M. R., Foster D. P., Godr\`eche C., and Mukamel D.,
J. Stat. Phys., in press (1995).

[6] Speer E. R., in {\it Micro, Meso, and Macroscopic approaches in Physics},
(Fannes M., Maes C., and Verbeure A., eds.), Proceedings of the NATO
AR workshop {\it On three levels}, Leuven (July 1993).

[7] Krug J., Phys. Rev. Lett. {\bf 67}, 1882 (1991).

[8] Derrida B., Domany E., and Mukamel D., J. Stat. Phys. {\bf 69}, 667 (1992).

[9]  Derrida B., Evans M. R., Hakim V., and Pasquier V., J. Phys. A, {\bf 26},
1493 (1993).

[10] Sch\"utz G. and  Domany E., J. Stat. Phys. {\bf 72}, 277 (1993).

[11] Bouchaud J. P. and Georges A., Phys. Rep. {\bf 195}, 127 (1990).

[12] Haus J. W. and Kehr K. W., Phys. Rep. {\bf 150}, 263 (1987).

[13] Feller W., {\it An Introduction to Probability Theory and Its
Applications},
Wiley (1968).

[14] See for example
Grimmett G. R. and Stirzaker D. R., {\it Probability and Random Processes},
Oxford University Press (1992).

[15] Evans M. R., unpublished.

[16] Godr\`eche C. and Luck J. M., unpublished.

[17] Ferrari P.A. and Fontes L.R.G., Ann. Probab. {\bf 22},
820--832 (1994).
}

\vfill\break

\def\bond#1#2{\hbox{${<}#1,#2{>}$}}
\def\+{\raise0.2ex\hbox{$\scriptstyle+$}}

\vskip 14 pt
\noindent
{\bf  Appendix A}
\vskip 12 pt

In this Appendix we show that the toy model is the $\beta\to0$ limit of the
two-species exclusion process, in a sense that will be made precise shortly.
The argument implies that this will be true even if the time over
which the models are compared, and the system size, become infinite as the
limit is taken.

 We begin with a summary of the basic strategy.  Let us write $\eta(t)$ for
the configuration of the exclusion process at time $t$; $\eta(t)$ is
determined by the initial configuration $\eta(0)$ together with some
specification of the (random) transitions during the evolution. Similarly,
$X(\tau)=(j(\tau),k(\tau))$ will denote the position of the walker in the
toy model at time $\tau$, and is determined by $X(0)$ and the random steps
taken by the walker. Now to each configuration $X=(j,k)$ of the toy model
there corresponds a configuration $\Phi(X)$ of the exclusion process:
$\Phi(X)\equiv(-\cdots-\,0\cdots0\,+\cdots+)$, with $j$ $(-)$ particles
and $k$ $(+)$ particles.  We will show that $\eta(t)$ agrees, with a
probability which approaches 1 as $\beta\to0$, with the time-rescaled image
$\Phi(X(2\beta t))$ of the toy process, assuming (a)~that the processes
coincide initially, i.e., $\eta(0) = \Phi(X(0))$, and (b)~that the random
transitions in the two models correspond.  Achieving the latter is called
{\it coupling} the models.

It is convenient to realize the randomness in the exclusion process as
follows.  There are $3(N-1)+4$ distinct types of elementary transitions
which occur in this process: three types of exchange ($+$ with $-$, $+$
with $0$, or $0$ with $-$) for each nearest neighbor site pair, and entrance
or exit of a particle at the left or right boundary.  With each type of
transition we associate a ``Poisson alarm clock,'' which is in fact just a
random set of positive times, Poisson distributed with density equal to the
rate---$q$, $1$, $\alpha$, or $\beta$---for the transition.
The alarm clock ``rings'' at each time $t^*$ in this set, and the
transition occurs if it is permitted in the configuration $\eta(t^*)$ at
that time.  (Here and below we assume, for definiteness, that $\eta(t)$ and
$X(\tau)$ are left continuous, so that if a transition from configuration
$\zeta$ to configuration $\zeta'$ occurs at time $t$, then $\eta(t)=\zeta$
and $\eta(t\+)\equiv\lim_{s\searrow t}\eta(s)=\zeta'$.)  We will refer
to these random times generically as {\it transition times} and more
specifically as {\it entrance times}, {\it left exit times}, etc.;
as emphasized above they are in fact {\it potential} transition times.

To couple the two models we use the random times from the exclusion
process to determine the evolution of the toy model; this coupling
corresponds to the intuitive description in Section~2.  Specifically, if
$t^*$ is a left exit time, then $\tau^*=2\beta t^*$ is a time at which
$j$ can decrease in the toy model.  Suppose that $X(\tau^*)=(j,k)$.  There
are several possibilities for $X(\tau^*\+)=(j',k')$:
 \par\noindent {\it Case I: $j=0$.} Nothing happens in the toy
model: $(j',k')=(j,k)$.
 \par\noindent {\it Case II: $j=1$.} In this case, $(j',k')=(0,N)$.
 \par\noindent {\it Case III:~$j\ge2$.} The transition in the toy
model is determined by looking ahead of $t^*$ to the first transition
time $t^{**}>t^*$ for one of two types of transitions: either for the
entrance of a $+$ particle, in which case
$(j',k')=(j-1,k+1)$,
or for a $0\,-\;\to\;-\,0$ exchange on bond $\bond12$, in which case
$(j',k')=(j-1,k)$, if $k>0$,
and $(j',k')=(N,0)$, if $k=0$.
 \par\noindent These possibilities are summarized in Table~A.1.  Right
exit times correspond similarly to times of possible decrease of $k$ in
the toy model.  Since the relative probabilities of cases III.1 and
III.2 are $\alpha/(1+\alpha)$ and $1/(1+\alpha)$, this prescription
reproduces the transition rules (2-4).  Here it is important that
transitions in the toy model are governed by the alarm clocks, not by
the actual occurrence of transitions in the exclusion process, since the
latter depend on the history of that process and hence their true
probabilities are not known.

\topinsert
\def\0{{\setbox0\hbox{${-}$}\hbox to\wd0{\hfil$0$\hfil}}}
 \def\s{height4pt&\omit&&&&&&&&&&\cr}
 \centerline{\vbox{\tabskip1em\baselineskip10pt\offinterlineskip\halign
{\vrule#&\hfil#\hfil&\vrule#&\hfil#\hfil&\vrule#&\hfil#\hfil&\vrule#&
  \hfil#\hfil&\vrule#&\tabskip0pt\hfil#&#\hfil\tabskip1em&\vrule#\cr
\multispan{12}\hrulefill\cr\s
&&&Allowed transition at time&&&&Final state&&&&\cr\s
&$j$&&$t^{**}$ in exclusion process&&$k$&&$(j(\tau^*\+),k(\tau^*\+))$&
   &\multispan2Case&\cr
\s\multispan{12}\hrulefill\cr\s
&$j=0$&&---&&$N$&&$(0,N)$&&I&&\cr
\s\multispan{12}\hrulefill\cr\s
&$j=1$&&---&&---&&$(0,N)$&&II&&\cr
\s\multispan{12}\hrulefill\cr\s
&&&\ $\0\,{-}\cdots\ \to\ {+}\,{-}\cdots$&&---&&$(j-1,k+1)$&&III&.1&\cr
\s&&\multispan{10}\hrulefill\cr\s
&$j\ge2$&&&&$k>0$&&$(j-1,k)$&&III&.2.a&\cr
\s&&&\smash{$\vcenter{\hsize1.6truein\hfil$\0\,{-}\cdots\ \to\
{-}\,\0\cdots$\hfil}$}\ \ &\multispan{8}\hrulefill\cr\s
&&&&&$k=0$&&$(N,0)$&&III&.2.b&\cr
\s\multispan{12}\hrulefill\cr}}}
 \medskip
\centerline{Table A.1: Transitions in the toy model for which
$j$ may decrease.}
\centerline{The initial state is $(j,k)=(j(\tau^*),k(\tau^*))$}
\medskip
\endinsert

 Suppose now that we observe the two systems over a time interval $[0,T]$
of the toy model---that is, for $0\le\tau\le T$ or $0\le t\le T/(2\beta)$.
Let $B\subset[0,T]$ be the set of (bad) times $\tau$ at which
$\eta(\tau/(2\beta))\ne\Phi(X(\tau))$, i.e., at which the systems do not
coincide. $B$ is a random set since it depends on the
transition times.  Now we would not expect agreement at all times,
since after the exit of a particle the exclusion system must rearrange
itself before it will again agree with the toy configuration; we can thus
at most hope that $B$ have small Lebesgue measure $|B|$.  Moreover, we
cannot rule out completely the possibility that $|B|$ will be large, since
if there are many exits in the exclusion process before the necessary
rearrangements occur, the eventual configuration may not agree with that of
the toy model.  These considerations motivate the following notion of
near-agreement: for $\epsilon>0$ we say that the two systems are {\it
$\epsilon$-close} if the probability that $|B|\ge\epsilon$ is less than
$\epsilon$.

 Now we can state the main result of this appendix.  We consider a limiting
process in which $\beta$ approaches $0$ and admit the possibility that
$T=T(\beta)$ and $N=N(\beta)$ increase during the limiting process; in
particular, they may approach infinity, but not too rapidly.
 \smallskip\noindent
 {\bf Theorem:} {\sl Suppose that $\eta(0)=\Phi(X(0))$ and that $T(\beta)$
and $N(\beta)$ are monotonically increasing functions which satisfy
$\lim_{\beta\to0}\beta NT=\lim_{\beta\to0}\beta T^2=0$.  Then
for any $\epsilon>0$ the two systems are $\epsilon$-close for sufficiently
small $\beta$.}
 \medskip\noindent
 {\sl Proof:} For uniformity of notation it is convenient to assume that
$T(\beta)\to\infty$ as $\beta\to0$. This entails no loss of generality,
since we may if necessary replace $T(\beta)$ by $T'(\beta)>T(\beta)$ with
$T'(\beta)\to\infty$ and with $T'(\beta)$ still satisfying the given
hypotheses (this is the only place that monotonicity is used), and observe
that if the two systems are $\epsilon$-close for $T'(\beta)$ then they are
also $\epsilon$-close for $T(\beta)$.

 Let $t_1,t_2,\ldots$ be the exit times for the exclusion process, in
increasing order, and let $M$ be the number of these which occur between $0$
and $T/(2\beta)$.  $M$ is a random variable having Poisson distribution with
mean and hence variance $T$, and Chebyshev's inequality implies that the
probability that $M\le 2T$ is at least $1-T^{-1}$ and hence goes to one as
$\beta\to0$.  Let us denote by $E_1$ the event that $M\le2T$.

Now let $r$ denote the minimum of the rates $1$, $q$, and $\alpha$, and
define
 $$a=a(\beta)=r^{-1}\left(4N+\beta^{-1/2}\right).\eqno({\rm A}.1)$$
 Because $1\ll a\ll\beta^{-1}$,
a time interval of length $a$ is small on the $\tau$ scale but large on the
$t$ scale.  It is during such an interval that agreement between the two
processes will be restored after a particle exit---that is, during an
interval $[t_i,t_i+a]$.  Restoration is simplest to analyze if no additional
exit occurs during the interval; note that the probability that the next exit
time $t_{i+1}$ satisfies $t_{i+1}<t_i+a$ is $1-e^{-\beta a}\le\beta a$.  Thus
the probability of the event that $t_{i+1}\ge t_i+a$ for all $i$ satisfying
$1\le i\le 2T$ is at least $1-2T\beta a$, which by hypothesis goes to one as
$\beta\to0$.  Let us call this event $E_2$.  When both $E_1$ and $E_2$ occur,
all the relevant exit times differ by at least $a$.

Suppose now that the two systems agree at the exit time $t_i$, i.e.,
$\eta(t_i)=\Phi(X(\tau_i))$, where $\tau_i=2\beta t_i$.  If an exit in fact
occurs in the exclusion process at time $t_i$---for a left exit, in cases II
and III of Table~A.1---then agreement will not hold immediately thereafter.
For each possible $X(\tau_i)\equiv(j,k)$, however, and knowing whether $t_i$
is a left or right exit time, we may identify a $(j,k)$-{\it restoration
event}: one or more sequences of entrance and exchange times whose occurrence
during the interval $(t_i,t_i+a)$ will, if no exit time lies in this
interval, guarantee the restoration of agreement by time $t_i+a$.  We will
describe these restoration events below and prove that the probability that a
$(j,k)$-restoration event does not occur in a time interval $(t,t+a)$ is, for
small $\beta$, uniformly bounded by $2(4N\beta+\beta^{1/2})$.  Thus the
probability of the event that a $X(\tau_i)$-restoration event occurs in the
interval $(t_i,t_i+a)$ for all $i$ satisfying $1\le i\le 2T$ is at least
$1-4T(N\beta+\beta^{1/2})$, which by hypothesis goes to one as $\beta\to0$.
Let us call this event $E_3$.

We can now complete the proof.  Take $\beta$ so small that the probability
that all of $E_1$, $E_2$, and $E_3$ occur is greater than $1-\epsilon$, and
consider the evolution of the system when these three events do occur.
Since the two systems agree at time zero they must agree until the first
exit time $t_1$.  Since no exit, and an $X(\tau_1)$-restoration event, occur
during the interval $(t_1,t_1+a)$, agreement will be restored by time
$t_1+a$, and will persist until time $t_2$.  Continuing in this way, we see
that agreement will hold for all $t$ between $0$ and $T/(2\beta)$ except for
at most $2T$ time intervals of length $a$.  Thus the total time of
disagreement is at most $4\beta aT$ on the $\tau$ time scale.  Finally, we
take $\beta$ so small that $4\beta aT<\epsilon$.

It remains to consider the restoration events and estimate their
probabilities. In the calculation we will assume that all entrance and
exchange processes occur at rate $r$, since in so doing we can only
underestimate the probability of restoration events.  Again let
$t_i=\tau_i/(2\beta)$ be an exit time---to be definite, a left exit
time---and set $X(\tau_i)=(j,k)$.  We divide the discussion according to
the cases of Table A.1.
 \par\noindent
 {\it Case III: $j\ge2$.\/} Agreement will certainly be restored if two
sequences of transition times, corresponding to cases III.1 and III.2, both
occur in the interval: (1)~a left entrance time, then $+\,-\to-\,+$
exchange times for bonds $\bond12,\ldots,\bond{j-1}j$ in succession, then
$+\,0\to0\,+$ exchange times for bonds
$\bond{j}{j+1},\ldots,\bond{N-k-1}{N-k}$, and (2)~$0\,-\to-\,0$ exchange
times for bonds $\bond12,\ldots,\bond{j-1}j$ in succession, followed if
$k=0$ (and hence $j=N$) by a right entrance time.  Each of these sequences
requires at most $N+1$ transition times and hence has probability of
nonoccurrence at most the probability that $U<N+1$, where $U$ is a Poisson
random variable with expectation (and hence variance) $ra$; by Chebyshev's
inequality,
 $$\mathop{\rm Prob}[\,U<N+1\,] \le {ra\over(ra-(N+1))^2}
  \le{4N+\beta^{-1/2}\over\beta^{-1}}=4N\beta+\beta^{1/2}.
\eqno({\rm A}.2)$$
Thus the event that one or the other sequence does not
occur has probability at most $2(4N\beta+\beta^{1/2})$.
 \par\noindent
 {\it Case II: $j=1$.\/} Agreement will be restored when the system fills
with $N-k$ $+$ particles.  It is easy to enumerate the entrances and
exchanges required, but these may occur in many different orders, so that a
direct computation of the time needed for restoration appears difficult.
We adopt an alternative approach by comparing the model to a
well-understood system: a totally asymmetric one-species exclusion process
on an infinite lattice, in which particles interchange with holes to their
right at rate $r$.  We suppose that sites in the one-species system are
independently occupied or empty with probability $1/2$---this is a steady
state---and couple the two systems by assuming that particle-hole exchanges
on bonds $\bond12,\ldots,\bond{N-1}N$ can occur in the one-species model
only at $+\,0\to0\,+$ exchange times for the same bonds in the two-species
model, and that a $+\,0\to0\,+$ exchange on bond $\bond01$ in the
one-species model can occur only at a left entrance time for the
two-species model.  If we let $J(u)$ denote the number of particles which
actually enter on the left in the two-species model during the time
interval $(t_i,t_i+u)$, and $J'(u)$ the number of exchanges which actually
take place on bond $\bond01$ in the one-species model during this same time
interval, then it is easy to verify that $J'(u)\le J(u)$ as long as
$J(u)<N-k$.  From this it follows that the probability that $J(a)<N-k$,
i.e., that restoration does not take place during the time interval
$(t_i,t_i+a)$, is at most the probability that $J'(a)<N-k$.  But it is
shown in [17] that $J'(a)$ has expectation $ra/4$ and variance $V(a)$
satisfying $\lim_{a\to\infty}V(a)/ra=0$; thus again by Chebyshev's
inequality,
 $$\mathop{\rm Prob}[\,J(a)<N-k\,] \le {V(a)\over(ra/4-(N-k))^2}
   \le{ra\over\beta^{-1}}=  4N\beta+\beta^{1/2}\eqno({\rm A}.3)$$
 for $\beta$ sufficiently small that $V(a)\le ra$.

\vfill\break

\noindent {\bf Appendix B}
\vskip 12 pt

This Appendix is devoted to an algorithm which allows for a fast recursive
solution of the difference equations~(4.10).

\noindent $\bullet$
We start with the $\v$ for the sake of simplicity.
Setting $j+k=n$, and introducing the notation $\v=\vv$,
the recursion relation~(4.10b) reads
$$
D\,\vv=F_j^{(n-1)}
,\eqno({\rm B}.1)
$$
with the boundary conditions $v_0^{(n)}=1$, $v_n^{(n)}=0$.
In eq.~(B.1) $D$ is the linear operator defined as
$$
D\,\vv=\vv-b\left(v_{j-1}^{(n)}+v_{j+1}^{(n)}\right)
,\eqno({\rm B}.2)
$$
and the source term reads
$$
F_j^{(n-1)}=a\left(v_j^{(n-1)}+v_{j-1}^{(n-1)}\right)
.\eqno({\rm B}.3)
$$

Eq.~(B.1) is a linear equation for the $v_j^{(n)}$,
with a source term $F_j^{(n-1)}$ involving the $v_k^{(n-1)}$,
whence the possibility of a recursive solution.

The solution to eq.~(B.1) is the superposition
$$
\vv=\left(v_{\rm H}\right)_j^{(n)}+\left(v_{\rm I}\right)_j^{(n)}
,\eqno({\rm B}.4)
$$
of the solutions of the following two equations
$$
\eqalignno{
D\, \left(v_{\rm H}\right)_j^{(n)}=0\hfill
&\quad\hbox{with}\quad \left(v_{\rm H}\right)_0^{(n)}=1,\quad
\left(v_{\rm H}\right)_n^{(n)}=0,&({\rm B}.5{\rm a})\cr
D\, \left(v_{\rm I}\right)_j^{(n)}=F_j^{(n-1)}
&\quad\hbox {with}\quad \left(v_{\rm I}\right)_0^{(n)}=0,\quad
\left(v_{\rm I}\right)_n^{(n)}=0.&({\rm B}.5{\rm b})\cr
}
$$
(with H for homogeneous, I for inhomogeneous).

The solution of the homogeneous equation~(B.5a) goes as follows.
Searching a solution of the form $z^j$, we find $z=\exp(\pm\sigma)$,
with the notation~(6.12).
Hence, imposing the boundary conditions, we obtain
$$
\left(v_{\rm H}\right)_j^{(n)}={U_{n-j}\over U_n}
,\eqno({\rm B}.6)
$$
where
$$
U_n(\zeta)={\sinh(n\sigma)\over\sinh\sigma}
\eqno({\rm B}.7)
$$
is the Chebyshev polynomial of the second kind and of order $n$ in the variable
$$
\zeta=2\cosh\sigma=1/b.
\eqno({\rm B}.8)
$$
These polynomials obey the recursion relation
$$
U_{n+1}+U_{n-1}=\zeta U_n
,\eqno({\rm B}.9)
$$
with $U_0(\zeta)=0$, $U_1(\zeta)=1$, so that $U_2(\zeta)=\zeta$,
$U_3(\zeta)=\zeta^2-1$, etc.

The solution of the inhomogeneous equation~(B.5b) reads
$$
\left(v_{\rm I}\right)_j^{(n)}=
\sum_{\ell=1}^{n-1}G_{j,\ell}^{(n)}\,F_{\ell}^{(n-1)}
,\eqno({\rm B}.10)
$$
where $G_{j,\ell}^{(n)}$ is the Green's function of $D$, defined by
$$
D\, G_{j,\ell}^{(n)}=\delta_{j,\ell}
\quad\hbox{with}\quad G_{0,\ell}^{(n)}=G_{n,\ell}^{(n)}=0
.\eqno({\rm B}.11)
$$
The Green's function can be obtained explicitly as a superposition
of plane waves $\exp(\pm j\sigma)$ in each of the sectors defined
by the inequalities $0\le j\le\ell$ and $\ell\le j\le n$.
We thus obtain the simple expression
$$
G_{j,\ell}={2\cosh\sigma\over U_n}\, U_\inf U_{n-\sup}
,\eqno({\rm B}.12)
$$
where $\inf$ denotes the smaller of the two integers,
i.e., $\ell$ if $\ell<j$ and $j$ otherwise,
whereas $\sup$ denotes the larger, i.e., $j$ if $\ell<j$ and $\ell$ otherwise.

The full recursive solution to eq.~(B.1) thus reads
$$
\vv={U_{n-j}\over U_n}+{\cosh\sigma-1\over U_n}\sum_{\ell=1}^{n-1}
U_\inf U_{n-\sup}\,\left(v_{\ell}^{(n-1)}+v_{\ell-1}^{(n-1)}\right)
.\eqno({\rm B}.13)
$$

\noindent $\bullet$
Likewise one may write a recursion relation for the $\uu$, namely
$$
\eqalign{
\uu&={2\cosh\sigma\over U_n}\sum_{\ell=1}^{n-1}U_\inf U_{n-\sup}\cr
&+{\cosh\sigma-1\over U_n}\sum_{\ell=1}^{n-1}
U_\inf U_{n-\sup}\,\left(u_{\ell}^{(n-1)}+u_{\ell-1}^{(n-1)}\right).
}
\eqno({\rm B}.14)
$$

One may deduce from the above analysis that the $\vv$ and $\uu$ are rational
expressions in $a$, or equivalently in $b$.
Their denominators are products of the Chebyshev polynomials
up to $U_n$.
As a consequence, as $n$ gets large,
the poles of the $\vv$ and $\uu$ become everywhere dense
in the range $-1<1/(2 b)<1$.
In terms of $b$, the poles cover the whole real axis
except the interval $[-1/2,\,1/2]$.
In terms of $a$, the poles cover the whole real axis
except the interval $[0,\,1]$.

The recursive solutions derived above
provide a fast and efficient way of computing the
$\u$ and the $\v$ numerically.

\noindent $\bullet$
The stationary probabilities $p_j^{(\nu)}$
can be determined by means of a similar recursive scheme.
First, eq.~(6.1) is a closed-form equation for the $p_j^{(0)}$,
whose solution reads
$$
p_j^{(0)}=p_0^{(0)}{U_j+U_{N-j}\over U_N},
\eqno({\rm B}.15)
$$
Second, the structure of eq.~(6.1) for $\nu\ge 1$
is very similar to that of eq.~(B.1).
We thus obtain the following recursion relation for the $p_j^{(\nu)}$
$$
p_j^{(\nu)}={\cosh\sigma-1\over U_n}\sum_{\ell=1}^{N-\nu-1}
U_\inf U_{N-\nu-\sup}\,\left(p_{\ell}^{(\nu-1)}+p_{\ell-1}^{(\nu-1)}\right).
\eqno({\rm B}.16)
$$
Third, the first probability $p_0^{(0)}$ is then determined
by means of the normalization condition (4.5).

\vfill\eject

\noindent {\bf Figure captions}
\vskip 12 pt
\parindent 0em

Fig. 1.
Characteristic curve $D(x,y)=0$ for $a=0.4$ (see eq.~(6.33)).
The dashed lines are the asymptotes.

Fig. 2.
Logarithmic plot of
$\Tt a^2$ as a function of $N a$ for the continuum limit (full curve)
and for the discrete random walk, with $a=0.1, 0.06, 0.03$.

\bye